\documentclass[preprint]{aastex}

\def\etal{{et~al.\,}}

\def\mj{$M_{\rm J}\,$}

\def\teff{T$_{\rm eff}\,$}
\def\teffs{T$_{\rm eff}$s$\ $}
\def\Dwa{$\,$\uppercase\expandafter{\romannumeral5}$\,$}
\def\mic{$\mu$m$\,$}
\def\undertext#1{$\underline{\smash{\hbox{#1}}}$}
\def\sles{\lower2pt\hbox{$\buildrel {\scriptstyle <}
   \over {\scriptstyle\sim}$}}
\def\under{\undertext}
\def\sgreat{\lower2pt\hbox{$\buildrel {\scriptstyle >}
   \over {\scriptstyle\sim}$}}
\def\sharpnull#1{}
\def\aa{Astron. Astrophys.\ }

\begin{document}

\slugcomment{\bf}
\slugcomment{Submitted to Ap.J.}

\title{Theoretical Spectral Models of T Dwarfs at Short Wavelengths and Their Comparison with Data}

\author{Adam Burrows\altaffilmark{1}, Adam J.\ Burgasser\altaffilmark{2}, 
J.\ Davy Kirkpatrick\altaffilmark{3}, James Liebert\altaffilmark{1},  
J. A. Milsom\altaffilmark{4},  D. Sudarsky\altaffilmark{1},  
and I. Hubeny\altaffilmark{5}}

\altaffiltext{1}{Department of Astronomy and Steward Observatory, 
                 The University of Arizona, Tucson, AZ \ 85721;
                 burrows@jupiter.as.arizona.edu, sudarsky@as.arizona.edu, liebert@as.arizona.edu}
\altaffiltext{2}{Division of Physics, M/S 103-33,
California Institute of Technology, Pasadena, CA 91125;
diver@its.caltech.edu}
\altaffiltext{3}{Infrared Processing and Analysis Center, M/S 100-22,
California Institute of Technology, Pasadena, CA 91125; davy@ipac.caltech.edu}
\altaffiltext{4}{Department of Physics, The University of Arizona, Tucson, AZ 85721;
                 milsom@physics.arizona.edu}
\altaffiltext{5}{NASA/GSFC, Code 681, Greenbelt, Maryland 20771;
                 hubeny@tlusty.gsfc.nasa.gov}

\begin{abstract}

We have generated new, self-consistent spectral and atmosphere 
models for the effective temperature range 600 K to 1300 K
thought to encompass the known T dwarfs.  
For the first time, theoretical models are compared with a {\it family} of measured T dwarf spectra 
at wavelengths shortward of $\sim$1.0 micron.
By defining spectral indices and standard colors in the optical and very near-infrared,
we explore the theoretical systematics with \teff, gravity, and metallicity.
We conclude that the short-wavelength range is rich in diagnostics that complement 
those in the near-infrared now used for spectral subtyping.  We also conclude that the wings 
of the Na D and K I (7700\AA) resonance lines and aggressive rainout of heavy metals
(with the resulting enhancement of the sodium and potassium abundances at altitude)
are required to fit the new data shortward of 1.0 \mic.  Furthermore, we find that 
the water bands weaken with increasing gravity, that modest decreases in
metallicity enhance the effect in the optical of the sodium and potassium lines, and that
at low \teffs, in a reversal of the normal pattern, optical spectra
become bluer with further decreases in \teff. 
Moreover, we conclude that T dwarf subtype 
is not a function of \teff alone, but that it is a non-trivial
function of gravity and metallicity as well.  
As do Marley et al. (2001), we see evidence
in early T dwarf atmospheres of a residual effect of clouds.
With cloudless models, we obtain spectral 
fits to the two late T dwarfs with known parallaxes,
but a residual effect of clouds on the emergent spectra 
of even late T dwarfs can not yet be discounted.
However, our focus is not on detailed fits to individual objects,
but on the interpretation of the overall spectral and color trends of the entire class of 
T dwarfs, as seen at shorter wavelengths. 

\end{abstract}

\keywords{infrared: stars --- stars: fundamental parameters ---
stars: low mass, brown dwarfs, T dwarfs, spectroscopy,
alkali metals, atmospheres, spectral synthesis}

\section{Introduction}
\label{intro}

The discovery of Gliese 229B ushered in a new chapter in stellar
astronomy by penetrating unambiguously below the main sequence edge 
(Nakajima et al. 1995; Oppenheimer et al. 1995).   
In Gliese 229B, absorption in the far infrared due to collision-induced absorption (CIA) by H$_2$
and the lack of absorption in the near infrared at the classic 
steam opacity windows together result in a redistribution of Gliese 229B's emergent flux 
from the mid- and far-IR to the $Z$ ($\sim$1.05 \mic), $J$ ($\sim$1.25 \mic), $H$ 
($\sim$1.6 \mic), and $K$ ($\sim$2.2 \mic) bands in the near-infrared.  These band fluxes exceed the corresponding
black body values for a given \teff by as much as two to three orders of magnitude.  Given the importance
of the near-infrared bands, it is only natural that they be
used to help establish the associated new spectral types (Burgasser et al. 2001a; Geballe et al. 2002).
Following on the heels of Gliese 229B, in the last few years 
more than twenty so-called ``T" dwarfs (Strauss et al. 1999; Burgasser et al. 1999,2000a,2000c,2001a; 
Cuby et al. 1999; Tsvetanov et al. 2000; Leggett et al. 1999,2000; Geballe et al. 2002)
and $\sim$200 so-called ``L" dwarfs (Ruiz, Leggett, and Allard 1997; Delfosse et al. 1997; 
Kirkpatrick et al. 1999,2000; Mart{\'{\i}}n et al. 1999) have been discovered 
(most by the 2MASS, SDSS, and DENIS surveys), introducing   
the first new ``stellar" spectral types since the 
establishment and articulation of the original MKK system (Morgan, Keenan, and Kellman 1943;
Morgan et al. 1992).  This new progression of M$\rightarrow$L$\rightarrow$T
from stars to brown dwarfs is one of the most exciting recent developments 
in ``stellar" astronomy.

While the use of the near-infrared to characterize and type L and T dwarfs
may be natural for low-temperature objects (most of which are 
substellar), this ignores their many interesting spectral features
and unique behavior shortward of 1.0 micron, the classic realm
of stellar classification.  It has recently been shown that the
potassium resonance doublet near 7700 \AA\ and the sodium D line(s)
around 5890 \AA , along with their broad wings, dominate T dwarf spectra
between 0.5 and 1.0 microns (Burrows, Marley, and Sharp 2000 (BMS); 
Tsuji, Ohnaka, and Aoki 1999; Liebert et al. 2000). It was also shown by BMS
that the I-band magnitude of Gliese 229B measured by 
Golimowski et al. (1998) and by Matthews et al. (1996)
is fully consistent with the potassium wing hypothesis.  
The resulting red/purple/magenta ``visual" appearance and the diagnostic (and at times,
counterintuitive; see \S\ref{system}) gravity, \teff, and metallicity dependences of
the short-wavelength spectra make the optical (loosely defined by the ``CCD" cutoff)  
from $\sim$0.4 \mic to $\sim$1.0 \mic an intriguing subject of study and theoretical inquiry.
Hence, in this paper we focus on the spectra of T dwarfs below 1.0 \mic.
We have calculated a new series of self-consistent 
spectra, colors, and spectral indices for theoretical brown dwarf atmospheres 
with \teffs\ from 600 K to 1300 K.  The low 
\teffs\ of these models and the fact that they do not incorporate
clouds/grains in their atmospheres exclude the L dwarfs from 
this inquiry (Burrows et al. 2001; Ackerman and Marley 2001).
However, although we concentrate here on the T dwarfs, we still find 
hints of the presence of clouds, particularly for the early Ts.   

In \S\ref{data}, we summarize some of the recent T dwarf measurements shortward
of 1.0 micron, which we interpret with the new theoretical models.  
In \S\ref{theory}, we present a full suite of new    
theoretical spectra.  These spectra will be used later in \S\ref{modelfits}--\S\ref{comparisons}
to extract information from the new T dwarf data at short wavelengths.  Section 
\ref{uncertain} summarizes the major techniques, assumptions, and uncertainties
in modeling of T dwarf spectra and includes an aside on the treatment
of the alkali line profiles. In \S\ref{profbright}, we give example  
T/P profiles and representative plots of the $\tau_{\lambda} = 2/3$ temperature, T$_{2/3}$.  Next, in 
\S\ref{define} we define the spectral indices and colors we use throughout, particularly
in \S\ref{comparisons}.  Note that \S\ref{profbright} and \S\ref{define}  
describe diagnostics by which one can better understand the observational trends. 
Then in \S\ref{modelfits}, we provide representative fits to the spectra of Gliese 229B and Gliese 570D 
and continue in \S\ref{rainout} with what can be learned 
with the new data at short wavelengths concerning 
the rainout of refractories (Burrows and Sharp 
1999; Lodders 1999) and the line profiles of the alkali metals.
After this, in \S\ref{system} we describe the systematic dependences of the optical and 
near-infrared spectra on \teff, gravity, and metallicity and in \S\ref{comparisons}
we compare the theoretical spectral indices with those obtained using T dwarf spectra.  
Finally, in \S\ref{clouds} we list several hints of the residual
influence of clouds in early T dwarf atmospheres (see also Marley et al. 2001)
and in \S\ref{conclusion} we summarize our general results and conclusions.
Given the remaining ambiguities in both the gas-phase opacities 
(e.g., of methane, the alkali metals, water) and in the proper treatment 
of clouds and given the possible effects of non-equilibrium chemistry 
(Griffith and Yelle 1999,2000; Saumon et al. 2000; Lodders 1999), we are
less interested in obtaining detailed fits than we are in the overall systematics
of the class of T dwarfs.  Hence, our focus in this paper is on the generic 
behavior at short wavelengths of the T dwarf {\it family} as a whole.

\section{The Observed T Dwarf Spectra Shortward of 1.0 \mic}
\label{data}

This is the first paper in which theoretical models and optical data for a large collection of T dwarfs,
not just individual T dwarfs such as Gliese 229B (Marley et al. 1996; Allard et al. 1996; 
Tsuji et al. 1996,1999; Saumon et al. 2001) or Gliese 570D (Geballe et al. 2001), are compared.
Table 1 itemizes 13 T dwarfs (and 2 late L dwarfs) with recently-obtained optical data. 
It lists objects in order of assigned spectral subtype (Kirkpatrick et al. 1999;
Burgasser et al. 2001a) and provides the associated references and short-hand names
we employ.  More detailed discussions on the Keck LRIS (Oke et al. 1995) optical
spectra of these T dwarfs and the associated signal-to-noise ratios 
can be found in Kirkpatrick et al. (2001) and Burgasser (2001a,b). 

Figures \ref{fig:1} and \ref{fig:2} depict spectra for the objects 
listed in Table 1 (except for 2MASS-1237) from 0.6 \mic to 1.0 \mic.
To facilitate intercomparison, these spectra have been normalized
to a value of one at 1.0 \mic.  Normalization also serves to emphasize that
we still don't know the parallaxes for most of these T dwarfs and, hence,
can not assign absolute flux levels.  The data shortward of $\sim$0.8 \mic
have a signal-to-noise ratio per pixel as low as $\sim$1 (Burgasser 2001b; Kirkpatrick et al. 2001).  Therefore,  
below 0.8 \mic we have used boxcar smoothing for the noisiest spectra to help discriminate 
between the relative flux levels of these T dwarfs in the $\sim$0.7 \mic region. 

As is clear from Figs. \ref{fig:1} and \ref{fig:2}, the spectral slopes
shortward of 1.0 \mic cover a broad range of values and the spectra themselves
fan out into a family that no doubt reflects variations in the underlying physical
properties.  Ignoring the two L dwarfs, the salient features are the water feature 
near 0.93 \mic, the spectral slope in the 0.8 \mic to 0.95 \mic region, and 
the relative height of the bump between the KI resonance lines ($\sim$7700\AA) and the Na D
lines ($\sim$5890 \AA).   These features inform our choice of the spectral indices we 
define for this study (\S\ref{define}).  Indeed, as was demonstrated by Liebert et al. (2000)
with their optical spectrum of SDSS-1624 (T6),
it is clear from Figs. \ref{fig:1} and \ref{fig:2} that the K I doublet at $\sim$7700\AA\ 
and its wings dominate the region from 0.7 \mic to 1.0 \mic (see \S\ref{theory}).
Also, as is clearest for the early T dwarfs and 2MASS-0559 (T5, green in Fig. \ref{fig:1}), absorption by the Na D 
lines is a natural explanation for the suppression of flux shortward of 0.7 \mic.
In principle, the equivalent widths of the Cs
features at 8523 \AA\ and 8946 \AA\ and of the Rb features at 7802 \AA\ and 7949 \AA\ are 
additional diagnostics (Basri et al. 1999; Griffith and Yelle 2000).
A useful index in this spectral region is the i$^{\prime}$-z$^{\prime}$ (Sloan AB; Fukugita et al. 1996) color,
where i$^{\prime}$ peaks in the region from $\sim$0.7 \mic to $\sim$0.8 \mic
and z$^{\prime}$ extends from approximately $\sim$0.8 \mic to $\sim$1.0 \mic.

The fact that the behavior of the spectra depicted in Figs. \ref{fig:1} and \ref{fig:2}
is not completely monotonic with the infrared-determined spectral subtypes 
indicates to us that the T dwarf spectral subtypes are not determined solely by \teff.   
Gravity and, perhaps, metallicity also play a role.  In particular,
the 2MASS-0559 (T5) (green curve in Fig. \ref{fig:1}) spectrum shortward of 0.9 \mic is ``redder"
than that of SDSS-1624 (T6) (blue curve in Fig. \ref{fig:1}), despite the former's earlier spectral subtype.
As we argue in \S\ref{comparisons}, this implies that the surface gravity of SDSS-1624 (T6)
is lower than that of 2MASS-0559 (T5). In addition, 
even though the spectral subtype of SDSS-1021 is T3, while that of SDSS-1254
is T2, the relative flux level of SDSS-1021 is generally above (at both 0.73 \mic
and 0.83 \mic) that of SDSS-1254.  Since in \S\ref{theory} and \S\ref{system}
we demonstrate in the likely \teff range of SDSS-1021 and SDSS-1254 
that decreases in \teff and increases in gravity increase the redness of the spectrum 
shortward of 0.9 \mic, we conclude that the gravity (and presumably the mass)
of SDSS-1021 is lower than that of SDSS-1254.   
Though \teff is the major determinant, a sufficiently large 
gravity difference can reverse the apparent dependence of subtype on changes in \teff.

The optical data depicted in Figs. \ref{fig:1} and \ref{fig:2} are state-of-the-art,
but, in particular for the later T dwarfs and shortward of 0.8 \mic, they are of fairly low resolution.  Nevertheless,
these spectra can help guide us in this exploration of the dependence of
optical spectra on T dwarf physical properties and in discerning the range 
of gravities and \teffs represented by the objects in Table 1.  The near absence
of parallaxes is an impediment to detailed fits, as is the noise in the observed spectra at
shorter optical wavelengths.  However, the optical colors 
and spectral indices (\S\ref{define} and \S\ref{comparisons})
that can be calculated for the growing list of T dwarfs with
measured spectra shortward of 1.0 \mic still provide 
useful physical diagnostics with which our suite of  
theoretical models can be used to determine the possible range 
of their \teffs, gravities, and metallicities.

\section{Theoretical Brown Dwarf Spectra from 1300 K to 600 K}
\label{theory}

As a prelude to the discussions in \S\ref{modelfits}--\S\ref{comparisons}, we present
a collection of new theoretical spectra that span the T dwarf regime.
Figure \ref{fig:3} portrays the absolute flux ($\cal{F}_\nu$) in milliJanskys versus wavelength
from 0.4 \mic to 1.5 \mic\ for solar-metallicity models for the two
gravities and the full range of \teffs (see \S\ref{uncertain}).  
The higher curves in Fig. \ref{fig:3} are for models with the higher \teffs. Prominent are
the Na D and K I resonance doublets at $\sim$5890 \AA\ and $\sim$7700 \AA, the water bands around 0.93 \mic,
1.15 \mic, and 1.4 \mic, the Cs I lines at 8523 \AA\ and 8946 \AA, the Li I line at
6708 \AA, the Rb I lines at 7802 \AA\ and 7949 \AA, and (for the hottest models) the TiO
and VO features near $\sim$0.45 \mic and 0.9$\rightarrow$1.05 \mic.
At 1.2432/1.2522 \mic the K I doublet is seen for \teffs at 700 K and above.
It was assumed in these models that clouds/grains/dust that may form at depth
nevertheless have no effect on an object's T/P profile, nor on its spectrum.
As we argue in \S\ref{clouds}, this may not be true for the earliest T dwarfs.
In determining molecular abundances, unless otherwise stated we employed the rainout
prescription for the refractory silicates described
in Burrows and Sharp (1999).  As explained in that reference
and in Burrows, Marley, and Sharp (2000), rainout depletes the atmospheres
of the refractory elements Ca, Al, Mg, Fe, and Si.  This affects the abundance
profiles of not only the nascent alkali metals, but TiO, VO, FeH, and CrH.
In particular, rainout suppresses the formation of alkali feldspars and
enables atomic Na and K to survive to lower temperatures and pressures, at which
point they form Na$_2$S(c) and KCl(c) (Lodders 1999).
As a consequence, their influence on the emergent spectrum in the ``optical" between 0.5 \mic
and 1.1 \mic is enhanced.  Rainout also restricts the range of lower temperatures and pressures
where TiO, VO, Fe(l,c), CrH, and FeH can be found.  However,
since the CrH opacities are soon to be updated significantly,
FeH and CrH bands were not incorporated into this model set and we defer
a discussion of their role to a later work (see also \S\ref{uncertain}).  

The normalized observed spectra shortward of 1.0 \mic were presented in Figs. \ref{fig:1} and \ref{fig:2}.
The corresponding normalized theoretical spectra at solar-metallicity
and for gravities of 10$^{5}$ cm s$^{-2}$ and 10$^{5.5}$ cm s$^{-2}$
are displayed in Fig. \ref{normal}.
A comparison of Figs. \ref{fig:1}, \ref{fig:2},
and \ref{normal} is illuminating and reveals that the theoretical family nicely spans the
observations.  Furthermore, the general overall spectral shapes, particularly from
0.8 \mic to 0.9 \mic, are reproduced.  We compare these model spectra with data 
in \S\ref{modelfits}-\S\ref{system}.  However, we first discuss
some of the major uncertainties in the models and a variety of physical diagnostics that are useful 
in interpreting measurements. 

\section{Model Assumptions, Techniques, and Uncertainties}
\label{uncertain}

To construct new atmosphere/spectral models of brown dwarfs,
we employ the complete linearization/accelerated-$\Lambda$-iteration
method of Hubeny and Lanz (1995), molecular and atomic opacities as
described in Burrows et al. (2001), and the equation of state of Saumon, Chabrier, and
Van Horn (1995). We explore both rainout, using the prescription of Burrows
and Sharp (1999) (see also Lodders 1999), and non-rainout composition assumptions,
though when not explicitly stated rainout has been incorporated.
The T/P profiles at depth obtained using the Lodders (1999)
and the Burrows and Sharp (1999) rainout prescriptions are the same (all else being equal)
to better than 2\% (M. Marley, private communication).

The depletion of refractories into silicate clouds is an intrinsically
non-equilibrium and dynamical process.  Hence, any prescription that
purports to address this can do so only crudely.  Fortunately, it is
the clouds at depth into which settle/rain the refractory 
elements that would have been in the upper atmospheres
that are the most problematically handled; the
upper atmospheric depletions of Mg, Si, Fe, Al, and Ca themselves and the resultant
metal-depleted molecular compositions are better handled, yielding 
compositions that are probably good to better than 10\%.
Since our modeling is focussed
on T dwarfs in which clouds play no central role (however, see \S\ref{clouds}),
our rainout prescriptions are as good as current practice allows.  However,
at the higher \teffs in the L dwarf regime, the actual physics of clouds (and their particle sizes,
spatial extent, optical properties, and degree of patchiness) 
assumes a greater importance. Furthermore, the possible effects
of non-equilibrium chemistry and rapid transport of spectroscopically-active
species may also be important (Lodders 1999; Griffith and Yelle 1999,2000; Saumon et al. 2000). 
In particular, though CO and NH$_3$ have no strong bands short of 1.0 micron
and play only a very minor role in setting the T/P profile of a T dwarf,
the suggestion that their abundances may be out of equilibrium due to dynamic convective transport
deserves further attention (Saumon et al. 2000; Noll, Geballe, and Marley 1997).
Quite obviously, no current prescription 
for cloud opacities or for dynamic transport and their effects on emergent
spectra can be beyond reproach or improvement. In this paper, given our 
focus on T dwarfs we ignore both.

Our models are produced using the opacity sampling technique to arrive
at converged and consistent temperature/pressure profiles, from which
higher-resolution spectra are then calculated for a given spectral interval.
For the T/P profile calculation, no more than 2000 wavelengths are
needed, which range from 0.4 \mic to 300 \mic.  Each model
was converged in temperature and flux to better than 0.1\%
in typically a few minutes on a standard workstation.  
The speed of our code is a direct consequence of the complete linearization approach.
Marley et al. (2001) and Burrows et al. (1997) use the k-coefficient 
method with the equivalent of $\sim$800 wavelengths.
Similar to the ODF method, their implementation avoids its major pitfall
by combining the abundance-weighted opacities of all the relevant
molecules and atoms before creating the distribution function that
is at the heart of the ODF approach.  We use a mixing-length prescription to
handle convection, with a mixing-length parameter of 1.0.  
Marley et al. (2001) flatten the entropy gradient in convective regions.  
Saumon et al. (2000) use the T/P profiles generated by Marley, as in Marley et al. (2001).
Allard et al. (2001) use mixing-length theory, with a mixing-length parameter of 1.0.
At low \teffs, the actual mixing length parameter is irrelevant; changing the
mixing-length parameter from 0.5 to 2.0 changes the T/P profile by no more than
$\sim$20 K.  However, above \teff $\sim$1500 K, the actual value of this parameter 
is an issue of modest interest.  
Nevertheless, the theoretical T/P profiles in grainless atmospheres below 
\teff = 1500 K are similar, differing by no more than 10\% 
(usually less) at atmospheric levels from 800 K to 2000 K 
(cf. this work; Marley, unpublished; Allard et al. 2001).

For the spectra we display here, 5000 wavelengths,
logarithmically spaced from 0.4 \mic to 1.5 \mic,
were used and then the spectrum was boxcar-smoothed to an effective
resolution, R ($\lambda/\Delta\lambda$), of 1000 at each wavelength.
Models were generated with \teffs from 600 K to 1300 K, in steps
of 100K (plus at a few extra \teffs, when needed), at two gravities
($g$ = $10^{5}$ cm s$^{-2}$ and $10^{5.5}$ cm s$^{-2}$), and for three metallicities (Z)
($0.3\times$solar, solar, and $2.0\times$solar). The Anders
and Grevesse (1989) abundance data were used to represent/define
the solar pattern.  This \teff-gravity-Z parameter set was chosen
to allow us to span the model space in which it is reasonable
to assume the newly-discovered T dwarfs reside, as well as to
gauge the systematic behavior of the observables with \teff,
gravity, and metallicity.

The brown dwarf radii ($R$) assumed
were derived from the analytic formula found in Burrows et al. (2001) and Marley et al. (1996):
\begin{eqnarray}
R &=& 6.7\times10^4{\rm km}\Bigl(\frac{10^5}{g}\Bigr)^{0.18}\Bigl(\frac{T_{\rm eff}}{10^3}\Bigr)^{0.11}\, ,
\nonumber
\end{eqnarray}
where $g$ is in cm s$^{-2}$ and \teff is in Kelvin. This procedure is not perfectly
consistent, since the evolutionary models upon which this equation for the radius is based
incorporate slightly different atmosphere models as boundary conditions.
Nevertheless, the error in the radius is small ($\sles$10\%) and the derived colors and spectral
indices are independent of radius.

The major differences between theoretical models arise from 
the different opacity databases employed.  For substellar dwarfs, the major gas-phase 
opacities are due to H$_2$O, H$_2$, CH$_4$, CO, NH$_3$, the neutral alkali metals,
TiO, VO, FeH, and CrH.  A subsidiary role is played by PH$_3$, MgH, CaOH, CaH, SiO, and H$_2$S.
Importantly, the {\it equilibrium} abundances of the dominant molecules (in particular,  
H$_2$O, H$_2$, CH$_4$, CO, N$_2$, and NH$_3$) 
are easily calculated (Burrows and Sharp 1999).
Most researchers use the same H$_2$ (Borysow and
Frommhold 1990; Zheng and Borysow 1995; Borysow, J\o{rgensen}, and Zheng 1997)
and H$_2$O (Partridge and Schwenke 1997) opacities. We calculate our TiO and VO opacities using Plez (1998,1999) and 
J\o{rgensen} (1997), including the effects of isotope shifts, and believe them to be state-of-the-art.  
As stated in \S\ref{theory}, we omit FeH and CrH from this T dwarf model set.  Their omission
affects predominantly L dwarfs in the spectral regions around 0.86 \mic
and 0.99 \mic (FeH, the Wing-Ford band), but does not change
to any significant degree T-dwarf T/P profiles, nor the vast majority of the
spectrum where FeH/CrH features do not contribute.
Note that, as do most researchers in this field (Saumon et al. 2000; Marley et al. 2001;
Allard et al. 2001), we currently use the 52-million line subset of the 308-million line Partridge and Schwenke (1997)
water database. Though recently questions have been raised concerning
its accuracy in some wavelength regimes (Jones et al. 2002), the Partridge
and Schwenke compilation remains the best available.

Methane opacities are an important wildcard, with the ``hot bands" and the red side
of the $H$ photometric band being the most problematic (Burrows et al. 2001).  However, apart from
a minor band near 0.89 \mic, methane plays only a very modest role in ``optical" T dwarf
spectra, though its effects in the infrared help to define the T dwarfs and the L$\rightarrow$T transition (Burgasser et al. 1999).  
The methane feature that has the most effect on a T/P profile is the strong $\nu_3$ band at 3.3 \mic,
but its opacity seems to be well in hand. 
Not unexpectedly, most researchers use much the same methane opacity data 
(cf. Burrows et al. 2001; Saumon et al. 2000; Marley et al. 2001).
Nevertheless, it is in the different implementations of the extant 
line lists and opacity databases, in the different collisional 
broadening prescriptions,  and in the inaccuracies in these various databases
that one is likely to find the origin of most of the differences between the various 
theoretical models (Burrows et al. 1997; Allard et al. 2001; Tsuji et al. 1999;
Saumon et al 2000; Geballe et al. 2001; Marley et al. 2001) and between theory and data.
However, the alkali lines and their wing profiles are a special case to which we now turn.

\subsection{An Aside on the Assumed Shapes of the Alkali Line Wings}
\label{alkali}

As emphasized in Burrows, Marley, and Sharp (2000), it is not
strictly correct to use Lorentzian
profiles for the Na D ($\sim$5890 \AA) and K I ($\sim$7700 \AA) doublets
in their far wings,  as we do for this paper and as is the standard practice in
most stellar atmospheres work. In fact, the dominance of these doublets
is a central feature of brown-dwarf/T-dwarf spectra and colors at wavelengths
from 0.4 \mic through 1.0 \mic.  While the line core behavior and oscillator strengths
for these transitions are well in hand (Piskunov et al. 1995), there remain ambiguities in the treatment
of these lines at large detunings ($\Delta\lambda$) and the as-yet-unknown shapes of their wings will have a bearing
on the viability of spectral fits.   For this study, we have used the default Lorentzian
out to a transition wavelength redward of the line cores (7700 \AA\ for Na D and 9800 \AA\
for the K I doublet), after which we allow the strength to decay as a Gaussian in wavelength
with a width of 0.075$\times\lambda_{\rm central}$.   The latter merely results in a smooth
cutoff at a large detuning, but is otherwise arbitrary.  This prescription for the alkali
wings is different from that suggested by Burrows, Marley, and Sharp (2000)
because in this paper we want to make only minimal alterations to the standard
Lorentzian until better estimates of the wing shapes are available.
The BMS prescription has an ad hoc parameter $q$ whose value affects the far wing
line shape and introduces an ``effective" cutoff, but is in fact unconstrained.
A choice of $q = 0.6$ for K and $q = 0.2$ for Na provides as good a fit
to T dwarf spectra as the simpler prescription we employ here, given the noise in the
optical data (Liebert et al. 2000).   However, the use of a BMS $q$ parameter implies
a certain functional form that is not motivated by the proper physical chemistry,
only by the need for a cutoff (\S\ref{rainout}).

Though the general overall spectral shapes seen in Fig. \ref{normal}, particularly from
0.8 \mic to 0.9 \mic, are reproduced (Liebert et al. 2000) with either the BMS  
or our modified Lorentzian prescriptions, there are two features that are
problematic in the comparision of Figs. 1, 2, and 4.  One is that the theoretical peak between the K I and Na D troughs
is shifted by $\sim$0.02 \mic to shorter wavelengths than the observed average.
This could easily be a consequence of the all-too-simple alkali line profile algorithms used.
The other is that for the earlier T dwarfs the K I trough
at 7700 \AA\ is a bit narrower than derived using
either the BMS or the current theory.  Both prescriptions yield equally discrepant spectral shapes,
though the fractional deviations in the flux densities or ``equivalent widths" derived using
either prescription are at most a few tens of percent.  
It should be noted that the prescriptions
for the alkali line shapes used in the literature have included a pure Lorentzian with a sharp cutoff at an arbitrary
wavelength at a large detuning (Allard et al. 2001), a pure Lorentzian with a sharp cutoff at an arbitrary
wavelength at a small detuning (Tsuji et al. 1999), and the Burrows, Marley, and Sharp (2000) prescription
(Marley et al. 2001; Saumon et al. 2000; Geballe et al. 2001).  Frequently, however, the alkali-line-shape prescription employed in a paper
is unexplained and sometimes the Na/K alkali metals have been omitted (Griffith, Yelle, and Marley 1998).

The data shortward of 0.8 \mic are indeed noisy, so this might be a major factor in any discrepancies
between our theoretical spectra and observation, but as the comparison of Figs. \ref{fig:1}, \ref{fig:2}, and \ref{normal}
nevetheless demonstrates, the remaining $\sim$10-50\% ambiguity in the flux densities is much smaller than
the factors of $\sim$2--10 deviations previously seen in the literature shortward of 0.8 \mic (cf. Tsuji et al. 1999).
Hence, our current approach to the Na/K alkali-line shapes is as reasonable as any,
but the reader should bear in mind the need for significant further improvement.

\section{T/P Profiles and the ``$\tau_{\lambda} = 2/3$" Temperature}
\label{profbright}

The dependence of the theoretical spectra on \teff, gravity, and metallicity
is a major focus of this paper.  Effects such as the general reddening in the optical with
decreasing \teff (at a given gravity), which reverses at lower effective temperatures
(particularly noticeable in Fig. \ref{normal} around 0.75 \mic), will be discussed in \S\ref{system}.
To facilitate that discussion, we present here representative model temperature/pressure
profiles in Fig. \ref{profiles} and representative plots of 
the T$_{2/3}$ temperature versus wavelength
in Fig. \ref{bright}.   T$_{2/3}$ is defined in this paper as the temperature
level in the brown dwarf atmosphere at which the total optical depth is $2/3$.
Roughly, it is the temperature of the layer to which one is probing when measuring
a spectrum at the corresponding wavelength and is a measure of the depth to 
which an observed spectrum is allowing us to peer (see also Saumon et al. 2000, their Figure 6).   
Fig. \ref{profiles} can be used to gauge the dependence  
on \teff, gravity , and metallicity of the optical depth above a given
temperature level in an atmosphere.  As Fig. \ref{profiles} suggests, 
lower-metallicity models have higher pressures at a given
temperature and higher-\teff models have lower pressures at a given temperature.
Moreover, as would be expected from hydrostatic equilibrium, higher-gravity 
models have higher pressures at a given temperature. 

T$_{2/3}$ is the ``decoupling" temperature at a given wavelength, or
the wavelength-dependent temperature level to which one probes an atmosphere by measuring its spectrum.
Fig. \ref{bright} indicates that shortward of 1.0 \mic, higher gravities result
in slightly lower T$_{2/3}$s and higher \teffs always result in higher
T$_{2/3}$s. In addition, Fig. \ref{bright} shows that in the $Z$ band ($\sim$1.05 \mic),
one is probing to $\sim$1500-1600 K for the 1100 K models and that in the $J$ band ($\sim$1.25 \mic),
one is probing to $\sim$1400-1500 K, for the same models.  Note that below the Na D line
at $\sim$5890 \AA,  T$_{2/3}$ for the sample models is rising 
to between 1200 and 1600 K, while at the centers of the strong Na D
and K I resonance doublets,  T$_{2/3}$ is near 800 K.

\section{Definitions of Color Indices}
\label{define}

Color-color diagrams are traditional tools used to determine
the physical properties of stars.  In addition, non-traditional,
but diagnostic, spectral indices (ratios of flux levels or flux averages
at different wavelengths) can perform the same function (Burgasser 
et al. 2000b; Burgasser 2001b; Geballe et al. 2002; Tokunaga and Kobayashi 1999).
Such indices are very useful for encapsulating and describing 
trends in \teff, gravity, and metallicity and in isolating
one class of objects from another.  This is especially true when spectral data are noisy,
when subtle changes in plotted spectra are difficult to discern by eye,
or when distances are not known (as is currently the case for the majority of T dwarfs).  
Hence, we have calculated a set of spectral indices and colors that
highlight various features of the family of T dwarf spectra, both
theoretical and observed, and use them to derive 
physical facts about the T dwarfs listed in Table 1, as well as  
about the spectral class as a whole. 
 
The indices we have defined are:
\begin{eqnarray}
X97 = \log_{10}(F_{\lambda}(0.90-0.91\mu {\rm m})/F_{\lambda}(0.72-0.73\mu {\rm m}))\, ,
\nonumber
\end{eqnarray}
\begin{eqnarray}
X98 = \log_{10}(F_{\lambda}(0.90-0.91\mu {\rm m})/F_{\lambda}(0.855-0.86\mu {\rm m}))\, ,
\nonumber
\end{eqnarray}
\begin{eqnarray}
X23 = \log_{10}(F_{\lambda}(0.92-0.925\mu {\rm m})/F_{\lambda}(0.928-0.945\mu {\rm m}))\, ,
\nonumber
\end{eqnarray}
and
\begin{eqnarray}
X126.105 = \log_{10}(F_{\lambda}(1.26\mu {\rm m})/F_{\lambda}(1.05\mu {\rm m}))\, .
\nonumber
\end{eqnarray}
Note that $F_{\lambda}$ (not $F_{\nu}$) is used, that a 
range of wavelengths implies an average in the indicated 
wavelength range, and that the indices do not have a 
prefactor of 2.5.  
These indices were chosen to highlight important features in T dwarf spectra
at shorter wavelengths. Both X97 and X98 were defined to avoid the water feature at $\sim$0.93 \mic.
X97 is a measure of the relative flux in the spectral bump
between the Na D and K I (7700 \AA) lines and the region around 0.9 \mic
and captures the effect of rainout. 
X98 is a measure of the slope between 0.8 \mic and 0.9 \mic
and, hence, is a measure of the shape of the red wing of the K I line at 7700\AA.  X23 is a measure
of the depth of the water feature near 0.93 \mic and reflects the
influences of H$_2$O abundance, gravity, and the residual effect of silicate clouds. X126.105 is a measure
of the ratio of the fluxes at 1.26 \mic and 1.05 \mic, near the traditional $J$ and $Z$
bands at the prominent peaks in T dwarf spectra (see Fig. \ref{fig:3}).
Our indices short of 1.0 \mic are similar to those
employed by Burgasser et al. (2001b) and Geballe et al. (2002), but are a
bit better tuned to capture physical effects, as opposed to
observational trends.  However, almost any of the ``optical" indices 
recently defined by those measuring L and T dwarfs could have 
sufficed.  (Tokunaga and Kobayashi (1999) do not define
indices short of 1.0 \mic, the spectral region on which 
we have focussed in this paper.)

We also use the the colors i$^{\prime}$ - z$^{\prime}$
and $J-K$, where the former is in the Sloan AB system
and the latter is in the Bessell (or Cousins) system (Bessell and Brett 1988).
Importantly, the choice of system
or index for the purpose of deriving trends is arbitrary.  We chose Bessell for the 
$J - K$ color to connect with the traditional color system in light of the continuing
confusion in the use of UKIRT, Ks, and 2MASS filters; there is as yet no uniformity  
from telescope to telescope in the use of near-infrared filter sets. 
Hence, our colors are just like spectral indices (e.g., X97 $\dots$) and 
those calculated and discussed here should be viewed as such.  
Our colors are derived from spectra, not photometry, and 
they are used to discover trends, not to compare with standard-star-calibrated values.  
Nevertheless, our i$^{\prime}$ - z$^{\prime}$ and $J - K$ (Bessell) colors 
are as good as any, given the differences between the underlying spectra of
calibration standards and T dwarfs and given the accuracy of the flux calibration
for the data (Burgasser et al. 2000a).  
Note that $J - K$ has been included in this paper, despite our emphasis 
on the optical, to make at least one tie-in with the spectral range now being
used to define the T dwarf spectral subtypes.

In Table 2, we provide these indices and colors for \teffs from 600 K to 1300 K, gravities
of 10$^{5}$ cm s$^{-2}$ and  10$^{5.5}$ cm s$^{-2}$, 
and metallicities of 0.3$\times$solar, solar, and 2.0$\times$solar, where solar
is defined by Anders and Grevesse (1989).  We include 
``no-rainout" models at 900 K and the two gravities and the observed T dwarfs listed in Table 1.
Table 2 summarizes all the basic trends (on which we focus in \S\ref{comparisons}). 
We use these indices in the following sections to facilitate the diagnosis 
of the observed spectral trends.

\section{Fits to Gliese 229B and Gliese 570D Spectra}
\label{modelfits}

That our prescription for the alkali line shapes has some merit is demonstrated in Fig. \ref{fittwo},
where solar-metallicity spectral fits for the two T dwarfs, Gliese 229B and Gliese 570D,
for which there are published parallaxes (Perryman et al. 1997; 5.8 and 5.9 parsecs, respectively)
are presented.  For Gliese 570D (as for the other T dwarfs in 
this paper), ours is the first paper to compare data with theory
below 0.8 \mic.  (For Gliese 229B, there are no spectral data as yet below 0.85 \mic, only
$R$ and $I$ band photometry.)  
The spectra portrayed in Fig. \ref{fittwo} are given in 
absolute flux units.  The shape of the spectrum from 1.0 \mic to
1.1 \mic, the slope of the spectra around 0.8 to 0.9 \mic, the Gliese 229B WFPC2 $R$
band measurement (Golimowski et al 1998), the flux level at $\sim$0.7 \mic for Gliese 570D, and the
i$^{\prime}$-z$^{\prime}$ color for Gliese 570D are all reproduced.
As Fig. \ref{fittwo} indicates, our theoretical fluxes  
for both Gliese 229B and Gliese 570D (which span more
than three orders of magnitude) are good to 10--30\%,
though the noise below 0.8 \mic is problematic.  Geballe et al. (2001) do not study
Gliese 570D shortward of 0.8 \mic.  However, they obtain comparably good
fits above 0.8 \mic, though we reproduce the shapes and 
magnitudes in the $Z$ and $J$ bands more closely.  Focussed on Gliese 229B and NH$_3$, 
Saumon et al. (2000) do not discuss spectra shortward of 1.0 \mic.  Marley et al. (2001) do not compare
theoretical spectra with observational data.  In their Gliese 229B campaign, 
Griffith, Marley, and Yelle (1998) and Griffith and Yelle (1999)
do not include the Na and K alkali lines in their opacity database.  Instead, they posit the existence at altitude
of a cloud of red grains whose imaginary index of refraction has a more extreme dependence on wavelength
than that of Titan tholins, red phorphorus, or polyacetylenes (Noy, Podolak, and Bar-Nun 1981; Khare and Sagan 1984).  
In their important Gliese 229B study, Tsuji et al. (1999) conclude
that a combination of silicate dust with the K I doublet at 7700 \AA\ can 
explain its spectrum shortward of 1.0 \mic.  However, as is still the standard practice in stellar atmospheres work, 
Tsuji et al. truncated the alkali lines at very small detunings, which necessitated the introduction of another 
component to explain the spectrum.  As did BMS, we find that including the wings of the potassium doublet
obviates the need for another component to explain the sharp declivity shortward of 1.0 \mic
in T dwarf spectra in general (Fig. \ref{fig:1} and \ref{fig:2}) and in Gliese 229B's spectrum in particular. 

As Fig. \ref{fittwo} indicates, the relative flux ratios in the $Z$ and $J$ bands
are well-modeled.  In contrast, the best-fit model of Gliese 229B by Tsuji et al. (1999) in the 
$Z$ and $J$ bands is off by a factor of 2--3, reflecting the need to cutoff the
alkali metal wings as discussed in \S\ref{alkali} and \S\ref{rainout}.  Importantly, the corresponding no-rainout
spectra below 0.8 \mic (not shown) (Allard et al. 2001) are generically off by a factor greater than 5 (\S\ref{rainout}).
Slightly more precise fits can in fact be achieved
by adjustments in \teff, metallicity, and gravity.  However, our purpose here is not to obtain the
ultimate fit, but to verify our general approach and to extract the essential conclusions about
the class of T dwarfs as a whole using spectra at short wavelengths.

Nevertheless, the depths of the water features near 0.93 \mic, 1.15 \mic
and 1.4 \mic (not shown) are not fit well.  Leakage in the spectrometer of light from
adjacent peaks into the troughs might explain the effect
at 1.4 \mic, and perhaps a fraction of the effect at 1.15 \mic, but
these are major discrepancies that may point to a residuum of dust in the atmosphere (Tsuji et al 1999; Saumon et al. 2000)
or to problems with the water opacity database used (Partridge and Schwenke 1997).  The models of all groups fail
to explain these water troughs (Saumon et al. 2000; Marley et al. 2001; Allard et al. 2001).  Lowering the metallicity enough,
particularly for a majority of the T dwarfs in Table 1 which all manifest
this problem, would not seem to be a viable option generically 
(Allard et al. 2001; however, see Griffith and Yelle 1999). 
Furthermore, a low-enough metallicity that would solve the problem
at 0.93 \mic would also shift the $J - K$ magnitudes away from the observed values
for the T dwarfs (Burgasser et al. 2001a) by as much as $-0.5$ magnitudes at 900 K.
(However, a lower metallicity may be implicated for 2MASS-0937; see \S\ref{comparisons}.)

A \teff for Gliese 570D of $\sim$750 K at a low gravity of 10$^{5}$ cm s$^{-2}$ (mass $\sim$30 \mj)
is similar to that derived in Geballe et al. (2001), but at their low end. Equally good
solar-metallicity fits to Gliese 570D can be found from $\sim$750 K to $\sim$840 K, where the corresponding gravities
range from $10^{5}$ cm s$^{-2}$ to $\sim$$10^{5.4}$ cm s$^{-2}$.
Our \teff range for Gliese 570D is a bit more conservative than that of Geballe et al. (2001) 
due to the fact that we have not imposed an age constraint.
However, as demonstrated in Fig. \ref{fittwo} for Gliese 229B, at the current level of observational precision
it is only a trajectory in \teff-gravity space (given roughly by \teff/$g^{0.2}$ = const.)
that can be constrained.  Gliese 229B could have a \teff near 780 K and a
gravity near 10$^{4.5}$ cm s$^{-2}$ (mass $\sgreat$15 \mj).  What is shown in Fig. \ref{fittwo} are just
representative fits to the Gliese 229B and Gliese 570D data.
Solar-metallicity models of Gliese 229B that fit these
spectra include those with \teffs from 780 K to 950 K (higher would imply a higher
gravity than is realistic for brown dwarfs).  Very subsolar-metallicity models for Gliese 229B
would compromise the acceptable fits in $J - K$.  

Without parallaxes, we prefer not to provide spectral fits for the 
other T dwarfs in Table 1.  It is not that fits can't be obtained.
That they can should be clear from Figs. \ref{fig:3} and \ref{normal}
and from Table 2.  Rather, the problem is that too many models can be found to fit,
and \teff and gravity are only loosely constrained.  Nevertheless, our models
and the indices discussed in \S\ref{comparisons} yield for all but
the early T dwarfs a \teff range of 900$\pm$150 K.  With parallaxes, we should 
be able to tighten this constraint and determine the \teff/$g$ trajectory
(and core entropy; Burrows, Marley, and Sharp 2000) for each observed T dwarf.  
However, at this stage, we don't want to claim more than is prudent.

\section{Conclusions Concerning Alkali Metal Line Profiles and Rainout}
\label{rainout}

There are a few conclusions of a qualitative nature, in particular concerning 
the truncation of the resonance lines of Na and K and the seeming necessity of rainout, that deserve
special mention.  Figure \ref{figcompare} demonstrates many of these
with a collection of representative theoretical model spectra  
at \teff = 900 K and solar metallicity from 0.4 \mic to 1.5 \mic. 
All the spectral models have T/P profiles that are consistently 
derived for the opacities and abundance prescriptions used.  
The solid red line depicts the [900 K/$10^{5}$ cm s$^{-2}$] model
and the solid blue line the [900 K/$10^{5.5}$ cm s$^{-2}$] model.  The black line
portrays a model with $g$ = $10^{5.5}$ cm s$^{-2}$, but uses Lorentzian line
profiles without cutoffs for the Na D and K I (7700 \AA) lines.  In addition, this spectrum does not
incorporate the rainout of Ti and V species, and, hence, leaves them suspended 
in the atmosphere in strict chemical equilibrium. The dashed blue line is a
[900 K/$10^{5.5}$ cm s$^{-2}$] model, but one that does not incorporate a prescription
for the rainout of the silicates and the consequent survival of Na and K to lower
temperatures and pressures in the atmosphere.  The spectrum of
2MASS-0559 (T5) is included on Fig. \ref{figcompare} in green and is the absolute flux spectrum under
the assumption that its distance is 10 parsecs. 

That unmodified Lorentzians produce spectra that do not fit is demonstrated
in Fig. \ref{figcompare} by the relative positions of the black line (Lorentzian model)
and the solid blue line.  The latter spectrum in the $Z$ ($\sim$1.05 \mic) and
$J$ ($\sim$1.25 \mic) bands is much closer to the data shown in Fig. \ref{fittwo}
and to those obtained for the objects in Table 1 by Burgasser (2001b);
the truncated models do not decapitate
the $Z$ and $J$ bands by as much as an order of 
magnitude in the way that the untruncated Lorentzian does.  As mentioned in \S\ref{modelfits}, the absence of such a cutoff is the
reason for the discrepancy by a factor of 2 to 3 in the Tsuji, Ohnaka, and Aoki (1999)
model for Gliese 229B in the $Z$ and $J$ bands.  The
difference at \teff = 900 K and $g$ = $10^{5.5}$ cm s$^{-2}$
between our standard model result and that with untruncated Lorentzians
is $\sim$1.2 magnitudes in i$^{\prime}$-z$^{\prime}$; the X98 and X97
indices are similarly incorrect (Table 2).

Note that, since Na is $\sim$20 times more abundant than K, it is Na
that would account for most of this anomalous suppression beyond 1.0 \mic,
despite the greater spectral distance of the Na D line. One concludes from 
Fig. \ref{figcompare} that the true
line profiles must be truncated and are not strictly Lorentzians.

As is clear from Fig. \ref{figcompare}, the no-rainout (dashed blue) and rainout (solid blue)
spectra differ by 0.5-1.0 dex from one another, particularly at 0.7 \mic 
between the dominating Na D and K I resonance doublets.  A comparison
of these curves with the new T dwarf spectra provided in Figs. \ref{fig:1} and \ref{fig:2}
leads to the strong conclusion that the no-rainout models do not fit the new optical data.  
This is clear confirmation of the BMS prediction (see their Figures 2 and 3)
for the ``entire" family of T dwarfs, not just for SDSS-1624 (Liebert et al. 2000),
and is consistent with the Marley et al. (2001) Figure 3.  Geballe et al. (2001) came
to a similar conclusion for Gliese 570D vis \`a vis rainout,
but did not have data nor theory shortward of 0.8 \mic.  Given the ambiguities
in the alkali line profiles near 1.0 \mic, spectra from 0.8 \mic to 1.0 \mic can not
be used to draw clear conclusions about the occurrence of rainout; it is only
with the new data at shorter wavelengths ($< 0.8$ \mic) that one can distinguish
unambiguously between the effects of rainout and of the unknown 
alkali line profile shapes, given the well-known
oscillator strengths.  

As Table 2 indicates, i$^{\prime}$ - z$^{\prime}$ for the no-rainout models
is as much as 1.0-1.5 magnitudes discrepant. The indices X98 and X97 tell
a similar story.  Without rainout, the ``redness" of the optical spectra 
and the large flux contrast between the $Z$ peak and the bump at 0.7-0.75 \mic
can not be reproduced.  Rainout leads to an enhancement in the abundance
of atomic Na and K at lower temperatures (and lower pressures) in the atmosphere
and to greater absorption shortward of 0.9 \mic.  A comparison on Fig. \ref{figcompare} of the spectrum
of 2MASS-0559 (T5) with the no-rainout/rainout theoretical curves serves to
emphasize this point (2MASS-0559 is here merely representative of the T dwarfs
listed in Table 1).  Hence, we have in the new T dwarf spectra at short
wavelengths evidence for the influence of cloud formation at depth and element
depletion (Si, Mg, Al) at altitude (Burrows and Sharp 1999; Burrows, Marley, and Sharp 2000;
Lodders 1999), leading to the suppression of alkali feldspar production at higher temperatures near $\sim$1400 K
and the consequent persistence of atomic Na and K to lower atmospheric temperatures.
In particular, the fit displayed in Fig. \ref{fittwo} for Gliese 570 D
and the approximate fit to the Gliese 229B datum in the $R$ band do not 
seem possible with any no-rainout models.  

As a test, for the Lorentzian model portrayed on Fig. \ref{figcompare} we suppressed
the effects of rainout on the TiO and VO abundances, thereby enhancing the abundances of those
molecules at altitude.  As this \teff = 900 K model demonstrates, around 0.45 \mic (near the $B$ band)
TiO features would be visible, even for a model with such a low \teff.   
This is a consequence of the low gaseous opacities at the shortest wavelengths.
The high T$_{2/3}$s (Fig. \ref{bright}) in this wavelength range say the same thing; 
at short wavelengths one is probing deeply.  If such features are not seen
(and this we suspect), it would be another indication of rainout.  Moreover, 
cloud veiling could also be implicated.  Hence, observations at wavelengths
even shorter than the Na D line(s) can provide information concerning the
deep atmosphere.

\section{The Systematic Dependence of T Dwarf Spectra on \teff, Gravity, and Metallicity}
\label{system}

From the spectra presented in Figs. \ref{fig:3}, \ref{normal}, and \ref{figcompare}, 
the representative T/P profiles and  T$_{2/3}$s shown in Figs. \ref{profiles} and \ref{bright}, and the
indices and colors tabulated in Table 2, we can determine the general theoretical trends for T dwarf
spectra with \teff, gravity, and metallicity.  

\subsection{Systematics with \teff}
\label{systeff}

Just as Fig. \ref{profiles} and all previous calculations indicate,
as the effective temperature decreases for a given gravity the pressure in the
atmosphere at a given temperature increases.  This implies that the column
depth in gm cm$^{-2}$ to a given temperature level increases with decreasing \teff 
(in hydrostatic equilibrium, column depth = $P/g$).  Hence, for a ubiquitous species such 
as water, the $\tau = 2/3$ surface that roughly determines the position of the ``photosphere" 
(on average, or at a given wavelength) is at progressively lower temperatures 
with decreasing \teff (see Fig. \ref{bright}). The result is that
T$_{2/3}$ in the water absorption
troughs near 0.93 \mic, and between the $Z$, $J$, and 
$K$ bands (for instance), as well as those at the emission peaks themselves, decreases.  However,
since the near-IR is not on the Rayleigh-Jeans tail,
these decreases in  T$_{2/3}$ lead to an increase in the flux
contrast from peak to valley, as well as to an overall decrease in the fluxes.
Hence, the water troughs deepen with decreasing \teff and, as Table 2 demonstrates,
$J - K$ gets {\it bluer} with decreasing \teff (Burrows et al. 1997; Allard et al. 2001). Roughly, this may  
be what is seen for the T dwarfs in the near IR (Burgasser et al. 2001a)  
and in Figs. \ref{fig:1} and \ref{fig:2}.   
At the higher \teffs (in the Ls and early to mid Ts), 
the systematics with decreasing \teff described above for water can also be seen
in the spectral region shortward of 1.0 micron dominated by the alkali metal lines
and wings.  In addition, increasing pressure with decreasing \teff leads 
to greater pressure-broadening, enhancing the influence of the alkali metal wings.
Both effects lead to a {\it reddening} of the spectrum below $\sim$1.0 \mic with decreasing \teff
which has been identified in previous theoretical explorations (e.g., BMS; Marley et al. 2001; Allard et al. 2001).

However, the Na and K abundances are very small below a certain temperature (1050 K to 850 K, depending on
element and pressure). Since there is such a ``top" to the atomic alkali region, 
as \teff decreases further at low \teffs the region containing Na and K atoms becomes more and more buried.
The result is a gradual diminution of the role of the Na and K lines from 0.4 to 1.0 \mic
and a consequent ``bluing" of the spectrum below $\sim$1.0 \mic.
This effect below 1.0 \mic was predicted by BMS; we now quantify it and  
in Fig. \ref{normal}, which shows the incipient closing of the K I absorption
trough at 7700 \AA\ and the slight bluing of the optical spectrum at low \teffs.
The gravity- and metallicity-dependent reversal of the indices X97, X98, and 
i$^{\prime}$-z$^{\prime}$ at low \teffs is captured in Table 2.  Recently, this prediction for 
the low \teff behavior of T dwarfs, as seen in i$^{\prime}$-z$^{\prime}$, has 
been verified by Marley et al. (2001).
The turnaround with decreasing \teff in the behavior of 
these spectral indices and of the i$^{\prime}$-z$^{\prime}$ color 
is even more clearly portrayed in Figs. \ref{fig:10}, \ref{fig:12}, and
\ref{fig:13} (discussed in \S\ref{comparisons}).  For the i$^{\prime}$-z$^{\prime}$ color, 
\teff for this transition ranges from $\sim$700 K to $\sim$800 K.  
In this \teff range, both i$^{\prime}$-z$^{\prime}$
and the overall spectral shape shortward of 1.0 micron are weak functions 
of \teff.

\subsection{Systematics with Gravity}
\label{gravity}

A T dwarf with a \teff of $\sles$1000 K and a gravity of 
10$^{5.5}$ cm s$^{-2}$ has a mass near 70 \mj 
(and an age $\sgreat$10$^{9.7}$ years).
One with a gravity of 10$^{5}$ cm s$^{-2}$ has a mass near 35 \mj.
As we go down in gravity to 10$^{4.5}$ cm s$^{-2}$,
we are exploring masses just above 15 \mj. The observed T dwarfs may possibly span
this entire range.  Figure \ref{profiles} shows that as the gravity
increases the pressure at a given temperature in the atmosphere 
increases as well.  What this figure does not show is that the column depth
of material actually decreases with increasing gravity.  This means
that for a given \teff, the depths of the water troughs and the contrasts
between the emission peaks and these troughs actually {\it decrease}
with increasing gravity.   In particular, we discover that the relative depth
of the water feature near 0.93 \mic (measured by X23) decreases with 
increasing gravity.  This is opposite to the effect of decreasing \teff.
One consequence is that any indices constructed from the contrasts
between peak and water trough are not a one-parameter family in \teff alone.
Hence, we find that the effect of decreasing 
\teff can be compensated by an increase in gravity. 
Importantly, as Table 2 demonstrates, increasing the gravity by 0.5 dex 
has the same effect on $J - K$ as decreasing \teff by 100-200 K.  
Hence, given the current method of T dwarf spectral subtyping 
in the near-IR (Burgasser et al. 2001a; Geballe et al. 2002),
effective temperature alone can not be construed to imply subtype.   At a given
\teff, lowering the gravity can lead to a later subtype.  Moreover, a later subtype
does not necessarily imply a lower \teff.     

As Table 2 indicates, $J - K$ gets bluer with increasing gravity.  This
is predominantly a consequence of the pressure dependence of the 
CIA opacity of H$_2$ at 2.2 \mic.  Shortward of 1.0 micron, at higher \teffs
increasing gravity reddens the i$^{\prime}$-z$^{\prime}$ color and increases (steepens)
the X97 and X98 indices.  However, at lower \teffs an increase in gravity
leads to a slight reversal and i$^{\prime}$-z$^{\prime}$ becomes bluer.  
This newly-quantified behavior can be seen in Fig. \ref{fig:12} (\S\ref{comparisons}) and Table 2
and is a consequence of the same systematics 
in the atmospheric profiles (Fig. \ref{profiles}) discussed earlier that 
are responsible for the reversal in the \teff dependence 
of the optical colors and slopes at lower \teffs.

\subsection{Systematics with Metallicity}
\label{smetal}

Figure \ref{figmetal} portrays the metallicity dependence of T dwarf spectra
from 0.4 \mic to 1.5 \mic, for a single representative \teff of 900 K and 
a gravity of 10$^{5.5}$ cm s$^{-2}$.  Metallicities at 0.3, 1.0, and 2.0 times
solar are compared.  Shortward of 0.9 \mic, we find that the 
lower-metallicity model is redder and the higher-metallicity model is bluer than
the solar model.  This newly-quantified effect may seem backwards, since one might have expected
a lower metallicity to weaken the effects of the alkali metals.
However, as the metallicity is lowered, the abundances of the other metals
(such as oxygen in the form of water) are also lowered, leading to a more transparent
atmosphere.  As a result, one is peering more deeply into the atmosphere
to higher pressures.  This is clearly apparent in the relative positions on
Fig. \ref{profiles} of the red (low-metallicity) lines. At higher pressures,
the Na D and K I resonance lines at $\sim$5890 \AA\ and $\sim$7700 \AA\ are stronger.  Hence, i$^{\prime}$-z$^{\prime}$
reddens with decreasing metallicity.   This is quite similar to what 
happens in subdwarf M stars.  Overall, the solar and 2.0$\times$solar
models are very similar shortward of 1.0 \mic. Unless the metallicity
is significantly subsolar (Griffith and Yelle 1999), the metallicity dependence of T dwarf spectra
in the optical is weak (except perhaps shortward of 0.6 \mic).
Note that our lower-metallicity model has a more rounded $Z$ band
peak. Such a shape is not in evidence in the T dwarf data. 

The differences in the near-IR at $H$ and $K$ are more
pronounced, with lower-metallicity models being generically more blue in $J - K$
and $H - K$.  For instance, from 600 to 1300 K, the differences between the
solar-metallicity and the 0.3$\times$solar-metallicity models vary from 0.1 to 0.5
magnitudes in $J - K$ (cf. Table 2 and Fig. \ref{fig:12}).  The higher pressures
in the low-metallicity atmospheres increase the CIA opacity and suppress
the $K$ band flux.  We find that decreasing the metallicity
by a factor of three has a greater effect on the $K$ band than
increasing the gravity by the same factor.  This behavior with metallicity
requires that if Gliese 229B has a low metallicity (Griffith and Yelle 1999) it must also have a
low gravity.  High-gravity models for Gliese 229B can not simultaneously 
be low-metallicity (Allard et al. 2001).  It would be very useful for the
determination of the physical parameters of the T dwarf Gliese 229B to obtain its spectrum
from $\sim$0.6 \mic to 0.8 \mic and its X97 and 
i$^{\prime}$-z$^{\prime}$ indices (consistently calibrated).

\section{Index and Color Comparisons}
\label{comparisons}

Figures \ref{fig:10}, \ref{fig:11}, \ref{fig:12}, and 
\ref{fig:13} are ``color-color" diagrams
that summarize the numerical data in Table 2. 
The upper-case letters on the figures represent the positions in these index plots 
of the measured T dwarfs.  The letter that corresponds to a given T dwarf
is indicated in Table 2 and in the figure caption to Fig. 
\ref{fig:10}.  Errors of $\sgreat$0.1 dex in the X indices and 
$\sim$0.25 in i$^{\prime}$-z$^{\prime}$ and $J - K$ (Bessell/Cousins) 
colors have been suppressed on these figures to avoid
further clutter and to allow the basic collective gestalt to emerge.    
The lines on the plots are families as a function of \teff at given gravities and metallicities.
Each point represents an \teff from 600 K to 1300 K, in steps of 100\ K.  
The basic success of the models is clear from the correspondence between the regions
occupied by the observed T dwarfs and where the theory points reside. In principle, a comparison 
of the observed positions on these plots with those for the theoretical models
would allow one to determine \teff, gravity, and metallicity.  
However, the errors in the indices and the overlaps of the model families make such
a determination a bit less straightforward.
In fact, such errors make it difficult to determine
\teff and gravity for most of the T dwarfs listed in Table 1 to better
than $\pm$150 K in \teff and 0.25 dex in gravity.  Fits to most of the measured
spectra can be obtained, but without parallaxes and given the remaining errors
in the data they tell us very little we can use.  Hence, we prefer not
to provide specific object-by-object \teff 
and gravity estimates, unless parallaxes are available or the object
is an outlier in its spectrum or indices.  Nevertheless, from Table 2,
Fig. \ref{normal}, the new optical data, and our theoretical models we can infer that the \teffs of the listed T dwarfs, 
except for the three early Ts, are 900$\pm$150 K.  The corresponding gravity range could 
in principle span our model set from 10$^{4.5}$ cm s$^{-2}$ to 10$^{5.5}$ cm s$^{-2}$.   
Note that values for the \teffs of the early T dwarfs are approximately bracketed by 
1100 K and 1300 K (see Table 2, Figs. \ref{fig:10},
\ref{fig:11},  \ref{fig:12}, and \ref{fig:13}).
Parallaxes would enable a determination of the possible \teff/$g$ line for
each object (as obtained in \S\ref{modelfits} for Gliese 229B) and 
are eagerly awaited.

From Fig. \ref{fig:10}, we see that our theory successfully provides the correct
general overall X97/X98 slope and distinguishes the L dwarf region from the T dwarf
region.  Furthermore, it demonstrates that if the observed early T dwarfs (green)
are not generically low-gravity objects (dashed), that the T dwarf edge has an 
effective temperature near 1300 K.  (If these early T dwarfs were low-gravity objects,
the T dwarf edge would be even lower in \teff.)  This L/T boundary temperature is 
in contrast with the $\sim$1600 K suggested by Basri et al. (1999).  For higher 
gravity, SDSS-1254 (T2,D) would have a \teff near 1200 K.
However, despite the inclusion of rainout, the models are still displaced downward
by $\sim$0.2 dex in X97. (Note that the non-rainout models [cf. Table 2] would be displaced
downward in X97 by a further $\sim$0.5 dex and to the left in X98 by $\sim$0.2 dex!)  

The fact that 2MASS-0937 (T6p,I), Gl 570D (T8,N), and 2MASS-0415 (T8,O) are
outliers is consistent with their lower \teff, higher gravity, or lower metallicity.
In particular, 2MASS-0937, due to its high X97 (and i$^{\prime}$-z$^{\prime}$; see Fig. \ref{fig:11})
and to the fact that in the near-IR it was typed by Burgasser et al. (2001a) as an earlier subtype than the other
two, seems clearly to have either a higher gravity or a lower metallicity
than the average T dwarf observed to date (Burgasser et al. 2001a). The lower-metallicity solution
is slightly favored when one looks at the other index figures and makes a consensus 
judgment. Furthermore, the $Z$-band spectrum of 2MASS-0937 (Burgasser et al. 2001a) is a bit more
rounded than average, reminiscent of the behavior of the low-metallicity model depicted 
in Fig. \ref{figmetal}.   We do not draw any conclusions from the lonely position of 2MASS-1237 (K)
on this figure and on Figs. \ref{fig:11} and \ref{fig:12} due to the noisiness 
of its spectrum between the Na D line and the K I line at 7700 \AA.  Better data are clearly
needed. 

Figure \ref{fig:11} demonstrates once again that our basic theoretical treatment 
shortward of 1.0 micron is not far wrong.  We also see that the \teffs of modest- to high-gravity
models of the early T dwarfs must be $\le$1300 K and that 2MASS-0937 (I) is best fit
with a low-metallicity model, though not definitively so.  However, the proximity 
of the model lines, their overlap, and the errors in the data together make it difficult to
draw detailed conclusions concerning the ``central" T dwarfs.  

Figure \ref{fig:12} is a color-color diagram that demonstrates the non-monotonic 
behavior of i$^{\prime}$-z$^{\prime}$ with \teff (\S\ref{system})
and the outlier status of 2MASS-0937.  It also makes the case that
the early T dwarfs are not fit well in the near infrared by this grainless
model set.  No doubt, part of the problem lies with the incomplete methane opacity
database in the $H$ and $K$ bands, notably the absence of 
the ``hot" bands of methane.  However, this can not account for the full
discrepancy.  The theoretical models are too blue in $J - K$ and the best
explanation for this is the influence of residual dust/grains/clouds even in the early T dwarfs
SDSS-0837 (T1,C), SDSS-1254 (T2,D), and SDSS-1021 (T3,E).   
The possible presence of grains is also suggested in Fig. \ref{fig:13}, 
which portrays both the behavior of X98 with X23 (H$_2$O jump)
and the non-monotonic behavior with \teff discussed previously.  Figure \ref{fig:13} shows
that the depth of the water feature near 0.93 \mic is not well reproduced
by the models, unless many of the observed T dwarfs are of low metallicity (Griffith and Yelle 1999,2000).  We surmise 
that extra opacity due to dust may be shallowing the water trough near 0.93 \mic by 0.1 to 0.15 dex.
Such dust may also be contributing to the general reddening shortward of 1.0 micron (Tsuji, Ohnaka, and Aoki 1999),
still dominated by the wings of the K I (7700 \AA) doublet.    

The index figures and Table 2 allow one to draw some qualitative conclusions
concerning specific T dwarfs. As indicated in Fig. \ref{fittwo}, Gliese 229B
can be fit by a low-g/low-\teff or a high-g/high-\teff model.  However, its  
value of X23 (J on Fig. \ref{fig:13}, one of the highest for the observed T dwarfs) implies 
that the low-gravity solution is preferred.  It is probable from their values
of i$^{\prime}$-z$^{\prime}$, X97, and X23 that SDSS-1624's gravity is lower than
that of 2MASS-0559.  Furthermore, a direct comparision of SDSS-1624's spectrum
with that of Gliese 229B leads one to conclude that its gravity is greater
than that of Gliese 229B.  In addition, a comparison of the i$^{\prime}$-z$^{\prime}$,
X98, and X97 indices for SDSS-1021 (T3) and SDSS-1254 (T2) leads one to conclude
that SDSS-1021 has the lower gravity.  This is also the conclusion arrived at 
in \S\ref{data} directly from the spectra.  SDSS-1021 could also have a higher \teff.

\section{Hints of Clouds}
\label{clouds}

The discussion in \S\ref{rainout} points to a quite interesting fact.
At \teffs near the early edge of the T dwarf family ($\sim$1300 K), T$_{2/3}$
in the $Z$ and $J$ bands reaches 1700-1800 K for atmospheres without the opacity effects
of clouds.  This suggests that the silicate/iron clouds
expected to reside above 1500-1700 K might significantly
affect the shape and strength of the $J$ and $Z$ emission peaks, even at such low \teffs.
A very similar conclusion was reached by Marley et al. (2001).
This is not unreasonable, since the L to T transition is thought to coincide
with the settling to depth of the clouds that dominate the
L dwarf spectral type (Ackerman and Marley 2001; Burrows et al. 2001). The spectra of the early T dwarfs
should show the residual influence of such clouds.  Indeed, as we argue in \S\ref{comparisons},
the (i$^{\prime}$-z$^{\prime}$)/($J - K$) color-color diagram suggests
that clouds are required to fit the early T dwarfs SDSS-0837 (T1),
SDSS-1254 (T2), and SDSS-1021 (T3).  As the T$_{2/3}$ temperature
argument suggests, it is in the $Z$ and $J$ bands where the effects of clouds are primarily expected.
Our spectral models at $\sim$1300 K (Fig. \ref{fig:3}) manifest TiO and VO features in these bands because
we have not included the effects of clouds.  Clouds should not only mute the
effects of TiO and VO on the $Z$ and $J$ bands of the \teff = 1200/1300 K models
in Fig. \ref{fig:3}, but should also help reduce the contrast between
the emission peaks at $Z$, $J$, and $K$ and the water troughs at
1.15 \mic and 1.4 \mic (Jones and Tsuji 1997).  Furthermore,
because the subordinate K I absorptions at 1.2432/1.2522 \mic
are from a thermally-excited state, they are formed only at high atmospheric
temperatures. Coincidently, they reside in the middle of the $J$
band where one is probing higher temperatures, as indicated by the higher
T$_{2/3}$s there (Fig. \ref{bright}).  However, the strengths we derive
for these subordinate lines (see Fig. \ref{fig:3}) are generally
greater than observed (McLean et al. 2000; Saumon et al. 2000).
This too may point to a mitigating role of veiling clouds/dust/grains.
Hence, there are numerous spectral regions that speak to the potential influence
of clouds on T dwarf spectra.

However, there remains a puzzle that may not be so easily explained by dust.
The classic Li feature at 6708 \AA\ is included in our model set
and can be seen on the curves in Fig. \ref{fig:3}.  Observations (Kirkpatrick et al. 1999)
suggest that the strength of this feature peaks in the mid-L spectral
range and then drops monotonically.  Currently, there is little indication of this
line in the observed late L dwarfs, nor in the T dwarfs.  However, in both our
rainout and no-rainout (unpublished) models the Li line survives
to below \teff = 600 K.  Allard et al. (2001) also obtain this feature
at low \teffs and Pavlenko (2001) has an extensive theoretical discussion
concerning its formation in late L dwarfs.  Since lithium burning in objects that currently
have T dwarf \teffs occurs only for a narrow range of masses
and gravities ($> 2.5\times 10^{5}$ cm s$^{-2}$), nuclear burning seems
to be excluded as a generic explanation.  Chemistry would be the natural
culprit, but both the Burrows and Sharp (1999) and Lodders (1999)
rainout prescriptions and the no-rainout prescription
(true equilibrium) give qualitatively the same
result.  Since atomic Li should exist in abundance to temperatures around 1300 K
(Lodders 1999) and should trail off below that
and since  T$_{2/3}$s around 6700 \AA\
are 1000 K to 1350 K for \teffs from 700 to 1300 K
(see Fig. \ref{bright} for 700$\rightarrow$1100 K),
there is no obvious reason the 6708 \AA\
Li line should disappear before the L/T boundary.
Therefore, we suspect there is an interesting story to be
told about real brown dwarf atmospheres
in the resolution of this Li-line problem.

\section{Conclusions}
\label{conclusion}

We have generated spectral models for the temperature range 600 K to 1300 K
thought to encompass the T dwarfs.  Comparing these models with the new 
T dwarf data shortward of 1.0 micron, we find that the models can explain
in qualitative and semi-quantitative fashion, the new observations and their systematic
trends. Furthermore, we demonstrate that untruncated Lorentzian line profiles
for the Na D and K I (7700 \AA) resonance doublets are disfavored, that 
the atmospheric abundances of sodium and potassium have indeed been enhanced by the rainout of
silicates (BMS; Lodders 1999), that water bands weaken with increasing gravity, that modest decreases in 
metallicity enhance the effect of the alkali lines in the optical, and that
at low \teffs the behavior of the optical spectra with \teff reverses and 
becomes bluer with further decreases in \teff (as predicted by BMS).  Moreover, we find that the 
upper edge of the T dwarf \teff range is near $\sim$1300 K.

We determine that the optical range is rich in diagnostics that are complementary
to those in the near-infrared now used for spectroscopic classification.  
An important conclusion is that the 
T dwarf subtype is not a function of \teff alone; subtype is also a non-trivial
function of gravity and metallicity. Even if the range of metallicities represented 
by the known T dwarfs is small, gravity will play an important role in the near-IR and shortward
of 1.0 micron in determining spectral shape, colors, and spectral subtype.   

From the shallowness of the water feature near 0.93 \mic, from the weaker than
predicted K I features in the $J$ band, from the failure of the models to fit the $J - K$
colors of the early T dwarfs, and from the presence in the hotter models
of TiO and VO features in the $Z$ and $J$ bands (not seen 
in the observations of the earlier T dwarfs), we find evidence 
of residual dust/grains/clouds in early T dwarf atmospheres.
A similar conclusion was recently reached by Marley et al. (2001).
The L dwarfs are known to be dominated by clouds
(Burrows et al. 2001, and references therein), and the transition from the L dwarfs to
the T dwarfs is predominantly due to the depletion and rainout of 
heavy metals (Ackerman and Marley 2001; Burrows and Sharp 1999;
Marley et al. 1996; Allard et al. 1996).  However, the continuing effect of clouds
in the early T dwarfs seems to be indicated by the data.  Clouds in T dwarf atmospheres
do not explain the redness of the spectrum shortward of 1.0 \mic (though they might contribute to it); this is 
mostly explained by the K I line at 7700 \AA\ and its wing (Burrows, Marley, and Sharp 2000).
However, the L and T transition is not as abrupt as earlier inferred by the
sharp swing to the blue of $J - K$.  One can still see deeply into the atmosphere,
even for low \teff, into regions occupied by silicate (and other?) clouds. 
The opacity spectrum of clouds is much more continuous and featureless
than that of gas.  As a consequence, the influence of clouds is manifest by  
broad shape changes in the optical and near-IR spectra and by the suppression of gas-phase spectral features,
not by specific and distinct spectral features, until one gets to $\sim$10 \mic
and the classical silicate band appears.  Certainly, good models of the particle
size spectrum, spatial extent, and composition of the grains or droplets
are required to reproduce the measured spectra of the early T dwarfs in detail.    
The later T dwarfs are easier to fit, but the residual effect of clouds at depth on 
the emergent spectra can not yet be discounted.  

The T dwarfs are all brown dwarfs and the number of such substellar objects known to astronomy is growing fast.
Their spectra hold the key to their physical characterization and shortward of 1.0 micron
these spectra have great diagnostic potential.  We have attempted to make
only a preliminary stab at comparing the new data at short wavelengths with theory and
look forward to the creation of better opacity databases and cloud models, 
more precise spectral measurements, and the measurement of parallaxes that together
will enable the final, detailed determination of the masses, gravities, \teffs, and compositions
of the objects that inhabit this fascinating new spectral class.

\acknowledgements

We are happy to thank Christopher Sharp, Curtis Cooper, Richard Freedman, Jonathan Lunine, Bill
Hubbard, Didier Saumon, Mark Marley, Hugh Harris, Conard Dahn, and Phil Pinto for insightful discussions and advice. 
For electronic versions of the synthetic spectra discussed in this paper, please contact the first author.
This work was supported in part by NASA grants NAG5-10760, NAG5-10629, NAG5-7499, and NAG5-7073. 
A.J.B. acknowledges support from the Space Telescope Science 
Institute through NASA/HST proposal \#8563. Portions of the data presented herein
were obtained at the W.M. Keck Observatory, which is operated as a scientific 
partnership among the California Institute of Technology, the University of
California, and the National Aeronautics and Space Administration.

\clearpage

 \begin{deluxetable}{cccc}
 \tablewidth{14cm}
 \tablenum{1}
 \tablecaption{T/L Dwarfs highlighted in this Paper}
 \tablehead{
 \colhead{$Full\ Object\ Designation$}&
 \colhead{$Short\ Object\ Name$}&\colhead{$Spectral\ Subtype$\tablenotemark{a}}&
\colhead{$Reference$}}
 \startdata
 2MASSI J1507038-151648           & 2MASS-1507 & L5& 1 \\
 2MASSW J1632291+190441           & 2MASS-1632 & L8& 1 \\
 SDSSp J083717.22$-$000018.3      & SDSS-0837 & T1& 2 \\
 SDSSp J125453.90$-$012247.4      & SDSS-1254 & T2& 2 \\
 SDSSp J102109.69$-$030420.1      & SDSS-1021 & T3& 2 \\
 2MASSI J0559191$-$140448         & 2MASS-0559 & T5& 3 \\
 SDSSp J162414.37+002915.6        & SDSS-1624 & T6& 4 \\
 2MASSI J0937347+293142           & 2MASS-0937 & T6p& 5 \\
 SDSSp J134646.45$-$003150.4      & SDSS-1346 & T6&  6 \\
 2MASSI J1237392+652615           & 2MASS-1237 & T6.5& 7 \\
 Gliese 229B                      & Gl 229B& T6.5 &8 \\
 2MASSI J0727182+171001           & 2MASS-0727 & T7& 5 \\
 2MASSI J1217110$-$031113         & 2MASS-1217 & T7.5& 5 \\
 2MASSW J1457150-212148           & Gl 570D & T8 &9 \\
 2MASSI J0415195$-$093506         & 2MASS-0415 & T8 &5 \\
\tablerefs{
(1) Kirkpatrick et al. (1999); (2) Leggett et al.\ (2000); (3) Burgasser et al.\ (2000c);
(4) Strauss et al.\ (1999); (5) Burgasser et al. (2001a); (6) Tsvetanov et al.\ (2000);
(7) Burgasser et al.\ (1999); (8) Nakajima et al.\ (1995); (9) Burgasser et al.\ (2000a).}
 \enddata
\tablenotetext{a}{Using the classification scheme of Kirkpatrick et al. (1999) for L dwarfs and
Burgasser et al. (2001a) for T dwarfs.}
 \end{deluxetable}

\clearpage

 \begin{deluxetable}{ccccccc}
 \tablewidth{12cm}
 \tablenum{2}
 \tablecaption{Spectral Indices for T/L Dwarfs and Z = (0.3,1.0,2.0)$\times{\odot}$ Models}
 \tablehead{
 \colhead{$Dwarf/Model$}&
 \colhead{$i^{\prime}-z^{\prime}$\tablenotemark{a}}&\colhead{$X97$\tablenotemark{b}}&\colhead{$X23$\tablenotemark{c}}&
 \colhead{$X98$\tablenotemark{d}}&\colhead{$J - K$\tablenotemark{e}}&\colhead{$X126.105$\tablenotemark{f}}}
 \startdata
2MASS-1507 (L5,A\tablenotemark{g}\ )                      &$  2.16$&$    0.68$&$    0.07$&$   -0.02$&$  -   $&$  -   $ \\
2MASS-1632 (L8,B)                      &$  2.88$&$    1.10$&$    0.17$&$    0.14$&$  -   $&$  -   $ \\
SDSS-0837 (T1,C)                      &$  3.61$&$    1.48$&$    0.13$&$    0.30$&$  1.07 $&$ 0.057   $ \\
SDSS-1254 (T2,D)                      &$  4.08$&$    1.58$&$    0.11$&$    0.37$&$  0.98 $&$ 0.120   $ \\
SDSS-1021 (T3,E)                      &$  3.92$&$    1.41$&$    0.11$&$    0.36$&$  0.84 $&$ 0.232   $ \\
2MASS-0559 (T5,F)                      &$  4.32$&$    1.69$&$    0.13$&$    0.38$&$  0.17 $&$ 0.188$ \\
SDSS-1624 (T6,G)                      &$  3.66$&$    1.34$&$    0.28$&$    0.37$&$  -   $&$  -   $ \\
2MASS-0937 (T6p,I)                     &$  4.99$&$    2.10$&$    0.35$&$    0.49$&$ -0.57 $&$   0.127$ \\
SDSS-1346 (T6,H)                      &$  4.36$&$    1.63$&$    0.44$&$    0.43$&$ -0.33 $&$   0.230$ \\
2MASS-1237 (T6.5,K)                    &$  3.86$&$    1.33$&$    0.28$&$    0.46$&$ -0.40 $&$   0.083$ \\
Gl 229B (T6.5,J)                 & $-$ & $-$ &$0.46$&$    0.52$&$    -0.16$&$  -0.300 $ \\
2MASS-0727 (T7,L)                      &$  4.06$&$    1.76$&$    0.29$&$    0.34$&$ -0.51 $&$   0.168$ \\
2MASS-1217 (T7.5,M)                   &$  3.83$&$    2.35$&$    0.34$&$    0.37$&$ -0.03 $&$   0.207$ \\
Gl 570D (T8,N)                   &$  4.38$&$    1.97$&$    0.53$&$    0.41$&$ -0.34 $&$   0.217$ \\
2MASS-0415 (T8,O)                      &$  4.01$&$    1.82$&$    0.35$&$    0.40$&$ -0.28 $&$   0.221$ \\
     &&&&&&\\

\hline
\hline
     &&&&&&\\
\under{Z = ${\odot}$}
     &&&&&&\\
     &&&&&&\\
g = $10^5$ cm s$^{-2}$:     &&&&&&\\
     &&&&&&\\
1300 K                      &$  3.10$&$    1.01$&$    0.27$&$    0.21$&$  0.48   $&$    0.32$ \\
1200                        &$  3.63$&$    1.29$&$    0.43$&$    0.33$&$  0.31   $&$    0.14$ \\
1100                        &$  3.91$&$    1.43$&$    0.53$&$    0.39$&$  0.23   $&$    0.13$ \\
1000                        &$  4.14$&$    1.52$&$    0.62$&$    0.40$&$  0.09   $&$    0.17$ \\
900                         &$  4.38$&$    1.64$&$    0.72$&$    0.42$&$  -0.05   $&$    0.22$ \\
800                         &$  4.53$&$    1.71$&$    0.81$&$    0.43$&$  -0.29   $&$    0.27$ \\
700                         &$  4.63$&$    1.78$&$    0.91$&$    0.45$&$  -0.53   $&$    0.35$ \\
600                         &$  4.51$&$    1.75$&$    0.99$&$    0.42$&$  -0.92   $&$    0.45$ \\
     &&&&&&\\
g = $10^{5.5}$ cm s$^{-2}$:     &&&&&&\\
     &&&&&&\\
1300 K                      &$  3.47$&$    1.19$&$    0.20$&$    0.29$&$  0.34   $&$    0.31$ \\
1200                        &$  3.83$&$    1.35$&$    0.28$&$    0.34$&$  0.18   $&$    0.34$ \\
1100                        &$  4.39$&$    1.60$&$    0.42$&$    0.45$&$  -0.09   $&$    0.16$ \\
1000                        &$  4.68$&$    1.73$&$    0.48$&$    0.48$&$  -0.24   $&$    0.20$ \\
900                         &$  4.74$&$    1.76$&$    0.52$&$    0.47$&$  -0.38   $&$    0.27$ \\
800                         &$  4.85$&$    1.82$&$    0.57$&$    0.48$&$  -0.69   $&$    0.35$ \\
700                         &$  4.70$&$    1.78$&$    0.63$&$    0.47$&$  -0.89   $&$    0.44$ \\
600                         &$  4.44$&$    1.72$&$    0.74$&$    0.42$&$  -1.26   $&$    0.50$ \\
     &&&&&&\\
\hline
     &&&&&&\\
No rainout:     &&&&&&\\
     &&&&&&\\
900 K/5.${\odot}$.norain                &$  3.20$&$    1.23$&$    0.83$&$    0.31$&$    -0.12$&$   -0.15$ \\
900/5.5${\odot}$.norain                 &$  3.08$&$    1.16$&$    0.64$&$    0.31$&$    -0.45$&$   -0.20$ \\
     &&&&&&\\
\hline
     &&&&&&\\
\under{Z = 0.3 ${\odot}$}
     &&&&&&\\
     &&&&&&\\
g = $10^5$ cm s$^{-2}$:     &&&&&&\\
     &&&&&&\\
1300 K             &$  3.19$&$    1.00$&$    0.17$&$    0.21$&$ 0.11    $&$    0.32$ \\
1200               &$  3.61$&$    1.21$&$    0.26$&$    0.28$&$ -0.09    $&$    0.25$ \\
1100               &$  4.16$&$    1.49$&$    0.36$&$    0.41$&$ -0.35    $&$    0.14$ \\
1000               &$  4.53$&$    1.65$&$    0.43$&$    0.46$&$ -0.48    $&$    0.13$ \\
900                &$  4.89$&$    1.80$&$    0.47$&$    0.49$&$ -0.71    $&$    0.16$ \\
800                &$  5.26$&$    1.97$&$    0.52$&$    0.53$&$ -1.01    $&$    0.19$ \\
700                &$  5.23$&$    1.97$&$    0.56$&$    0.54$&$ -1.27    $&$    0.25$ \\
600                &$  5.17$&$    1.97$&$    0.61$&$    0.54$&$ -1.82    $&$    0.30$ \\
     &&&&&&\\
g = $10^{5.5}$ cm s$^{-2}$:     &&&&&&\\
     &&&&&&\\
1300 K             &$  3.56$&$    1.17$&$    0.13$&$    0.28$&$ -0.04    $&$    0.20$ \\
1200               &$  4.15$&$    1.46$&$    0.18$&$    0.39$&$ -0.34    $&$    0.18$ \\
1100               &$  4.59$&$    1.64$&$    0.22$&$    0.46$&$ -0.55    $&$    0.23$ \\
1000               &$  4.89$&$    1.77$&$    0.26$&$    0.50$&$ -0.70    $&$    0.20$ \\
900                &$  5.23$&$    1.90$&$    0.29$&$    0.54$&$ -0.90    $&$    0.18$ \\
800                &$  5.44$&$    2.00$&$    0.32$&$    0.59$&$ -1.22    $&$    0.21$ \\
700                &$  5.29$&$    1.95$&$    0.35$&$    0.56$&$ -1.51    $&$    0.26$ \\
600                &$  5.11$&$    1.90$&$    0.40$&$    0.54$&$ -2.01    $&$    0.29$ \\
     &&&&&&\\
\hline
     &&&&&&\\
\under{Z = 2.0 ${\odot}$}
     &&&&&&\\
     &&&&&&\\
g = $10^5$ cm s$^{-2}$:     &&&&&&\\
     &&&&&&\\
1300 K             &$  3.20$&$    1.11$&$    0.42$&$    0.25$&$ 0.72    $&$    0.10$ \\
1200               &$  3.44$&$    1.20$&$    0.48$&$    0.26$&$ 0.58    $&$    0.21$ \\
1100               &$  3.89$&$    1.44$&$    0.66$&$    0.39$&$ 0.43    $&$    0.16$ \\
1000               &$  4.13$&$    1.55$&$    0.77$&$    0.39$&$ 0.29    $&$    0.11$ \\
900                &$  4.12$&$    1.55$&$    0.84$&$    0.38$&$ 0.30    $&$    0.18$ \\
800                &$  4.29$&$    1.65$&$    0.97$&$    0.39$&$ 0.17    $&$    0.26$ \\
700                &$  4.37$&$    1.71$&$    1.12$&$    0.40$&$ -0.05    $&$    0.35$ \\
600                &$  4.17$&$    1.66$&$    1.24$&$    0.34$&$ -0.36    $&$    0.45$ \\
     &&&&&&\\
g = $10^{5.5}$ cm s$^{-2}$:     &&&&&&\\
     &&&&&&\\
1300 K             &$  3.69$&$    1.30$&$    0.35$&$    0.36$&$  0.42    $&$    0.12$ \\
1200               &$  3.92$&$    1.41$&$    0.43$&$    0.40$&$  0.34    $&$    0.09$ \\
1100               &$  4.06$&$    1.48$&$    0.49$&$    0.41$&$  0.25    $&$    0.13$ \\
1000               &$  4.17$&$    1.53$&$    0.56$&$    0.41$&$  0.14    $&$    0.22$ \\
900                &$  4.41$&$    1.66$&$    0.65$&$    0.43$&$  0.06    $&$    0.29$ \\
800                &$  4.46$&$    1.70$&$    0.74$&$    0.44$&$  -0.12   $&$    0.38$ \\
700                &$  4.39$&$    1.69$&$    0.82$&$    0.42$&$  -0.43   $&$    0.48$ \\
600                &$  4.11$&$    1.61$&$    0.96$&$    0.36$&$  -0.73   $&$    0.56$ \\
     &&&&&&\\
\hline

 \enddata
\tablenotetext{a}{Sloan Digital Sky Survey (SDSS) AB band passes; 3631 Jy. zeropoints}
\tablenotetext{b}{X97 = $\log_{10}(\langle F_{\lambda}(0.90-0.91\, \mu{\rm m})\rangle/\langle F_{\lambda}(0.72-0.73\, \mu{\rm m})\rangle)$}
\tablenotetext{c}{X23 = $\log_{10}(\langle F_{\lambda}(0.92-0.925\, \mu{\rm m})\rangle/\langle F_{\lambda}(0.928-0.945\, \mu{\rm m})\rangle)$}
\tablenotetext{d}{X98 = $\log_{10}(\langle F_{\lambda}(0.90-0.91\, \mu{\rm m})\rangle/\langle F_{\lambda}(0.855-0.86\, \mu{\rm m})\rangle)$}
\tablenotetext{e}{$J - K$ (Bessell); $J$ zeropoint: 1600 Jy., $K$ zeropoint: 655 Jy.}
\tablenotetext{f}{X126.105 = $\log_{10}(F_{\lambda}(1.26\, \mu{\rm m})/F_{\lambda}(1.05\, \mu{\rm m}))$}
\tablenotetext{g}{Letter used to distinguish position of T dwarf on Figs. \ref{fig:10},
\ref{fig:11}, \ref{fig:12}, and \ref{fig:13}.} 
 \end{deluxetable}
\clearpage

\voffset=+0.1pc  

\figcaption{
The log (base ten) of the flux ({\cal F}$_\nu$) versus wavelength 
($\lambda$) in microns from 0.6 \mic to 1.0 \mic for two late L dwarfs
(2MASS-1507 and 2MASS-1632) and some representative T dwarfs 
for which optical data have recently been obtained (see Table 1
for references).  All spectra have been normalized
to be 1 milliJansky at 1.0 \mic so that the relative flux levels are more 
easily compared.  The objects are listed in the graph from early to late
spectral subtype (indicated next to the object name).  Due to the 
low signal-to-noise ratio shortward of 0.8 \mic for the 
latest T dwarfs and the desire to more easily discriminate objects,
boxcar smoothing of from 10 to 20 \AA\ has been applied below 0.8 \mic for 2MASS-1624,
2MASS-0937, and Gliese 570D.  Note that the 2MASS-0559 (T5) (green) spectrum
shortward of 0.9 \mic is ``redder" than that of SDSS-1624 (T6) (blue), despite
the former's earlier spectral subtype. 
\label{fig:1}}

\figcaption{
The log (base ten) of the flux ({\cal F}$_\nu$) versus wavelength
($\lambda$) in microns from 0.6 \mic to 1.0 \mic for six additional T dwarfs.  This
figure is a continuation of Fig. \ref{fig:1} and on it the data have
been similarly normalized.  Furthermore, boxcar smoothing, as described
in Fig. \ref{fig:1}, has been applied to SDSS-1346, 2MASS-0727, 2MASS-1217,
and 2MASS-0415.  Both this figure and Fig. \ref{fig:1} include Gliese
229 B (in gold) to enable cross comparison.  Since SDSS-1021 and SDSS-1254
are rather close on this figure, an arrow has been used distinguish SDSS-1021.
Note that the relative flux level of SDSS-1021 (T3) is generally above (at both 0.73 \mic
and 0.83 \mic) that of SDSS-1254, despite the former's later spectral subtype.
\label{fig:2}}

\figcaption{
The log (base ten) of the absolute flux ({\cal F}$_\nu$) in milliJanskys versus wavelength
($\lambda$) in microns from 0.4 \mic to 1.5 \mic for self-consistent 
theoretical solar-metallicity (Anders and Grevesse 1989) models 
of brown-dwarf/T-dwarf spectra generated for this paper.  These spectra
have been deresolved to an $R$($\lambda/{\Delta\lambda}$) of 1000.  
The dashed blue lines are for a gravity of $10^{5.5}$ cm s$^{-2}$ 
and the red lines are for a gravity of $10^{5}$ cm s$^{-2}$.
Shown are models from 600 K to 1300 K, in steps of 100 K.  The higher lines
are for models with the higher \teffs. Prominent are
the Na D and K I resonance doublets at $\sim$5890 \AA\ and $\sim$7700 \AA, the water features around 0.93 \mic,
1.15 \mic, and 1.4 \mic, the Cs I lines at 8523 \AA\ and 8946 \AA, the Li I line at
6708 \AA (for which see \S\ref{clouds}), the Rb I lines at 7802 \AA\ and 7949 \AA, and the TiO
and VO features near $\sim$0.45 \mic and 0.9$\rightarrow$1.05 \mic (for the hottest
models).  No clouds (which would affect the detectability of the TiO/VO features, 
among others) are incorporated into these models. Note that the K I
doublet at 1.2432/1.2522 \mic is seen for \teffs at 700 K and above.
FeH and CrH bands are not incorporated into this 
model set (the corresponding opacity data are being
updated for future publication).
\label{fig:3}}

\figcaption{
The same set of solar-metallicity theoretical models depicted in Fig. \ref{fig:3}, but over
a wavelength range from 0.6 \mic to 1.0 \mic and normalized at 1.0 \mic
to a universal value of 1.0 (as in Figs. \ref{fig:1} and \ref{fig:2}).  The dashed blue lines are
for a gravity of $10^{5.5}$ cm s$^{-2}$ and the red lines are for a gravity of $10^{5}$ cm s$^{-2}$.
This figure is meant to be compared with Figs. \ref{fig:1} and \ref{fig:2}.
The general reddening trend in the optical with
decreasing \teff (at a given gravity) reverses at lower effective temperatures.
At higher gravities, this reversal occurs at higher \teffs (for a given metallicity).
This effect is particularly noticeable around 0.75 \mic .
See text for a discussion.
\label{normal}}

\figcaption{
Atmospheric profiles of the temperature (T, in K) versus the logarithm (base ten) of the pressure (in dynes cm$^{-2}$)
for different metallicities and for \teffs of 900 K (solid) and 1300 K (dashed, one dotted).
A gravity of 10$^{5.5}$ cm s$^{-2}$ was used for the solid and dashed models,
while the lower-gravity model at 1300 K and twice solar metallicity
is depicted with a dotted line.  Note that pressure is plotted
upside down, with the lower pressures at the top.  The 2.0$\times$solar-metallicity models
are blue, the solar-metallicity models are black, and the 0.3$\times$solar-metallicity
models are red.  Lower-metallicity models have higher pressures at a given
temperature and higher-\teff models have lower pressures at a given temperature.
Lower-gravity models have lower pressures at a given temperature, as a comparison
between the blue dashed and dotted curves demonstrates.
Temperature/pressure profiles help determine the character of the emergent spectrum.
\label{profiles}}

\figcaption{
The ``$\tau_{\lambda}$" temperature (T$_{2/3}$) in Kelvin versus the wavelength ($\lambda$) in microns
for \teffs of 700, 900, and 1300 K and gravities of $10^{5.5}$ cm s$^{-2}$ (thick, blue)
and $10^{5}$ cm s$^{-2}$ (red) for the solar-metallicity models shown in
Fig. \ref{fig:3}.   T$_{2/3}$ is defined in this paper as the temperature
level in the brown dwarf atmosphere at which the total optical depth is $2/3$.
Hence, it is a measure of the ``decoupling" temperature at a given wavelength, or
the wavelength-dependent depth to which one probes an atmosphere. Note that,
very roughly, the wavelengths at the flux peaks (see Fig. \ref{fig:3}) are
where  T$_{2/3}$ is highest.  For instance, in the $z^{\prime}$ band ($\sim$1.05 \mic),
one is probing to $\sim$1500-1600 K for the 1100 K models and in the $J$ band ($\sim$1.25 \mic),
one is probing to $\sim$1400-1500 K, for the same models.  Note that below the Na D line
at $\sim$5890 \AA,  T$_{2/3}$ is rising to between 1200 and 1600 K, while at the centers of the strong Na D
and K I resonance doublets, T$_{2/3}$ is near 800 K.  See the text for a discussion of the
conceptual usefulness of T$_{2/3}$.
\label{bright}}

\figcaption{
A comparison of the absolute fluxes (F$_{\nu}$), in milliJanskys) of Gliese 229B (gold) and Gliese 570D (red)
with representative solar-metallicity model spectra for wavelengths ($\lambda$) from 0.6 \mic to 1.4 \mic.
Gliese 229B should be compared with the models at [950 K/10$^{5.5}$ cm s$^{-2}$] (upper black)
and [780 K/10$^{4.5}$ cm s$^{-2}$] (green) and Gliese 570D should be compared with
the model at [750 K/10$^{5}$ cm s$^{-2}$] (lower black).  The gold bars near 0.65 \mic denote the
WFPC2 $R$ band measurement of Gliese 229B of Golimowski \etal (1998).  The fits are good,
but, at the current precision, not unique.
\label{fittwo}}

\figcaption{
The log (base ten) of the absolute flux ({\cal F}$_\nu$) in milliJanskys versus wavelength
($\lambda$) in microns from 0.4 \mic to 1.5 \mic for various \teff = 900 K
solar-metallicity models.  The red line depicts the [900 K/$10^{5}$ cm s$^{-2}$] model
and the solid blue line the [900 K/$10^{5.5}$ cm s$^{-2}$] model.  The black line
portrays a model with $g$ = $10^{5.5}$ cm s$^{-2}$, but using strict Lorentzian line
profiles for the Na D and K I (7700 \AA) lines, as well as no rainout for the
Ti/V compounds (note the region around 0.45 \mic). The dashed blue line is a
[900 K/$10^{5.5}$ cm s$^{-2}$] model, but one that does not incorporate a prescription
for the rainout of the silicates and the consequent survival of Na and K to lower
temperatures and pressures in the atmosphere.  For comparision, the spectrum of
2MASS-0559 (T5) is included in green and is the absolute flux spectrum under
the assumption that its distance is 10 parsecs. The rainout and no-rainout models
(solid blue versus dashed blue) are very different shortward of 1.0 \mic. Even for the low \teff (900 K)
models shown here, the no-rainout model can not even qualitatively reproduce the
steeper spectral slope and lower bump between 0.65 \mic and 0.75 \mic seen
generically in the optical/near-IR data (see Figs. \ref{fig:1} and \ref{fig:2}),
as exemplified here with the 2MASS-0559 (T5) spectrum.
See text for discussion.
\label{figcompare}}

\figcaption{
This is a plot of the log (base ten) of the flux (F$_{\nu}$) in milliJanskys versus
the wavelength ($\lambda$) in microns from 0.4 \mic to 1.5 \mic of model spectra
at \teff = 900 K and with a gravity of 10$^{5.5}$ cm s$^{-2}$ for metallicities at 0.3,
1.0, and 2.0 times solar (Anders and Grevesse 1989).  Shortward of 0.9 \mic, the
lower-metallicity model is redder and the higher-metallicity model is bluer than 
the solar model.  Furthermore, the lower-metallicity model has a more rounded $Z$ band
peak. Overall, the 1.0 and 2.0$\times$solar models are very similar.  However, 
there are metallicity differences shortward of 0.6 \mic, as well as in the water troughs
(particularly around 0.93 \mic). The differences in the near-IR at $H$ and $K$ are more
pronounced, with lower-metallicity models being generically more blue in $J - K$
and $H - K$.  For instance, from 600 to 1300 K, the differences between the 
solar-metallicity and the 0.3$\times$solar-metallicity models vary from 0.1 to 0.5 
magnitudes in $J - K$.
\label{figmetal}}

\figcaption{
Index X97 ($\log_{10}(F_{\lambda}(0.90-0.91\mu{\rm m})/F_{\lambda}(0.72-0.73\mu{\rm m}))$)
versus index X98 ($\log_{10}(F_{\lambda}(0.90-0.91\mu{\rm m})/F_{\lambda}(0.855-0.86\mu{\rm m}))$) 
for a set of theoretical models and 
a subset of the observed T and late L dwarfs listed in Table 1.
X97 is a measure of the relative fluxes at $\sim$0.9 \mic and the bump between
the K I resonance doublet at 7700 \AA\ and the Na D line(s) 
(see Figs. \ref{fig:1}, \ref{fig:2}, \ref{fig:3}, and \ref{normal}).
X98 is one measure of the spectral slope between (but not including) the K I doublet and the water feature
at $\sim$0.93 \mic.
These indices are also given in Table 2.
This and the following ``pseudo" color-color diagrams help to quantify the behavior 
at ``short" wavelengths of that part of a brown dwarf spectrum
that is dominated by the resonance alkali metal lines.
For the theory, the connected black dots and lines are for solar metallicity,
those in gold are for 0.3$\times$solar metallicity, and those in blue are
for 2.0$\times$solar metallicity.  The dashed lines are for a gravity of 
$10^{5}$ cm s$^{-2}$ and the solid lines are for a higher gravity of $10^{5.5}$ cm s$^{-2}$.  
Each theory line connects models from 1300 K to 600 K, in steps of 100 K.
Notice the hook in these lines are lower \teffs, even though much of the 
higher \teff behavior is monotonic. The basic trends and
systematics can be gleaned at a glance, though extracting
the specifics requires closer scrutiny.  The single letters
denote observed objects: the blue letters are the L dwarfs 2MASS-1507 (A) and
2MASS-1632 (B), the green letters are SDSS-0837 (C), SDSS-1254 (D),
and SDSS-1021 (E), and the red letters are the later T dwarfs, 2MASS-0559 (F),
SDSS-1624 (G), SDSS-1346 (H), 2MASS-0937 (I), Gl 229B (J, not shown here),
2MASS-1237 (K), 2MASS-0727 (L), 2MASS-1217 (M), Gl 570D (N), and 2MASS-0415 (O).
They are in order of spectral subtype, as derived by Burgasser et al. (2001a) (see Table 2).
Refer to the text for a general discussion
of this and related Figures \ref{fig:11} through \ref{fig:13}.
Note that any error bars for the indices have been suppressed for clarity, but they are generally
around 0.1 dex.
\label{fig:10}}

\figcaption{
The same as Fig. \ref{fig:10}, but for i$^{\prime}$-z$^{\prime}$ (Sloan) versus X98. 
The basic trends in the theory lines are clear.
(See text for a discussion and Table 2 for a listing.)  
As stated in Fig. \ref{fig:10}, error bars for the data points
have been suppressed for clarity, but they are $\sim$0.1 dex for X98 and $\sim$0.2$^{+}$ mag
for i$^{\prime}$-z$^{\prime}$.
\label{fig:11}}

\figcaption{
The same as Fig. \ref{fig:10}, but for i$^{\prime}$-z$^{\prime}$ (Sloan) versus $J - K$ (Bessell).
The use of the Bessell (Cousins) system in this plot is arbitrary and the quoted colors have
been derived from the spectra themselves. As used, our $J - K$ (Bessell) value can be considered merely 
yet another index to gauge relative slopes and shapes.  Note that fluxing and absolute calibration
for the spectra of the depicted T dwarfs may have errors as high as 0.25 mag.  In addition,  
note that the difference between the photometrically-derived (not used) and 
spectroscopically-derived magnitudes can at times be of a similar magnitude (Burgasser et al. 2001a).  
The theoretical and observed spectra have been treated equally in obtaining the numbers (see Table 2) for this
plot.  On this plot, H (SDSS-1346) and N (Gl 570D) are very close (see Table 2). 
\label{fig:12}}

\figcaption{
The same as Fig. \ref{fig:10}, but for X98 versus X23 ($\log_{10}(F_{\lambda}(0.92-0.925\mu{\rm m})/F_{\lambda}(0.928-0.945\mu{\rm m}))$).
Note that, unlike on Figs. \ref{fig:10} through \ref{fig:12}, Gliese 229B (J, gold)
is found on this plot.  X23 is a measure of the relative depth of the water feature
at $\sim$0.93 \mic.
\label{fig:13}}

\end{document}